\documentclass[%
 reprint,
 amsmath,amssymb,
 aps,
 pre,
]{revtex4-2}

\usepackage{graphicx}% Include figure files
\usepackage[rgb]{xcolor}
\usepackage{dcolumn}% Align table columns on decimal point
\usepackage{bm}% bold math
\usepackage{mathtools}
\usepackage{hyperref}% add hypertext capabilities
\definecolor{MyBlue}{RGB}{0,0,220}
\definecolor{MyRed}{RGB}{214,39,40}

\begin{document}

%\preprint{APS/123-QED}

%\title{Common features of and differences between models for active particles}
\title{Models for active particles: common features and differences}

\author{Colin-Marius Koch}
\author{Michael Wilczek}%
 \email{michael.wilczek@uni-bayreuth.de}
\affiliation{%
 Theoretical Physics I, Universität Bayreuth, Universitätsstraße 30, 95447 Bayreuth
}%

\date{\today}

\begin{abstract}
Systems of active particles can show a large variety of collective behavior. In theory, two aspects determine the collective behavior: the model at the particle level and the parameter regime. While many studies consider a single model and study its parameter regime, here, we focus on the former aspect. Motivated by experiments that study dilute suspensions of Chlamydomonas reinhardtii in a self-generated oxygen gradient, we compare various models with external field-dependent motility to understand how the collective behavior changes between models. We vary the particle-particle interaction from no interactions to steric interactions, the particle shape from round disks to dumbbells, the self-propulsion mechanism from constant speed to rocking motion, and the particle's center of mass from the geometric center to off-center. We find that changes in the model of the active agents can lead to similar statistics in the dilute regime and different collective behavior in the dense regime. We conclude that active particle models do not easily generalize for different real active agents, but instead require a clear understanding of the agents' microscopic properties. 
\end{abstract}

\maketitle

Currently, there exist a plethora of models for systems of active particles~\cite{romanczuk2012active,bechinger2016active,zottl2023modeling}. The models describe the particles' properties, such as shape~\cite{moran2022shape,moran2022particle,hopkins2023motility}~and propulsion mechanism~\cite{lauga2009hydrodynamics,moran2017phoretic,shaik2023confined}, as well as the interaction with other particles and the surrounding, such as mutual repulsion~\cite{fily2012athermal,buttinoni2013dynamical,sanoria2021influence} and motion along chemical gradients~\cite{schnitzer1993theory,liebchen2018synthetic,shaik2023confined}. Each model, in turn, features parameter regimes that distinguish, e.g., between persistent and diffusive motion~\cite{deseigne2010collective,cates2015motility} or dense and dilute systems~\cite{digregorio2018full}. Both, the model and the parameter regime, can have a significant impact on the collective behavior of active particles~\cite{cates2015motility}. When trying to describe real active agents, it is often unclear which model details are necessary to match the observed behavior. It may occur that multiple models produce the same statistics in one regime, but show different collective behavior in another regime. This study aims to stimulate interest in the discussion of how generalizable active particle models are.

A recent work studied the collective behavior of Chlamydomonas reinhardtii inside a self-generated oxygen gradient~\cite{fragkopoulos2021self}. Under low light conditions, C. reinhardtii consume oxygen for self-propulsion and deplete the local oxygen concentration. The authors observed an aggregation of C. reinhardtii in the center of a circular, quasi-two-dimensional chamber when oxygen could diffuse inwards only through the chamber's boundaries. In contrast to earlier works on particles in external fields~\cite{schnitzer1993theory,cates2015motility,stenhammar2016light,frangipane2018dynamic,arlt2019dynamics}, the authors measured a relationship between particle speed $v$ and density $\rho$ that is $v\propto\rho^{-0.5}$ instead of $v\propto\rho^{-1}$. So far, it remains unclear which microscopic details of the C. reinhardtii are responsible for the observed discrepancy to earlier literature. While we do not aspire to model this experiment in all its details, it motivated us to abstract the system and study the impact of microscopic properties on the statistics and collective behavior.

\begin{figure}
\includegraphics[width=\columnwidth]{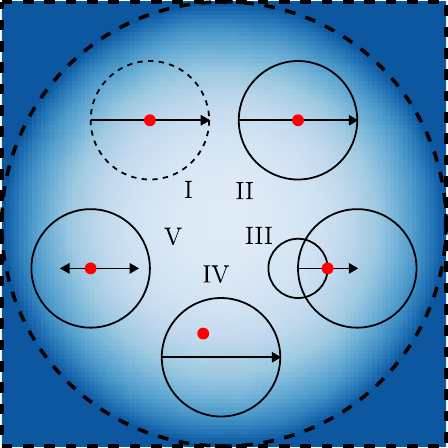}
\caption{Five models for active particles inside an external field (blue color code). The particles can fully move through the square box with periodic boundary conditions and deplete the external field inside the dashed circular area. Solid lines indicate excluded volume interactions and the red dots mark the particles' centers of mass.}
\label{fig:models-sketch}
\end{figure}

So, in contrast to studies that explore the parameter regimes of a single model, here, we compare five different models for active particles, see Fig.~\ref{fig:models-sketch}. For all models, the particles' motility depends on an external field that is consumed by the particles~\cite{fragkopoulos2021self}. Models I and II feature non-interacting as well as sterically repulsive active disks~\cite{fily2012athermal,digregorio2018full}, respectively. Model III considers active dumbbells~\cite{wensink2014controlling,wysocki2015giant,ostapenko2018curvature,cammann2021emergent} and model IV active disks with a shifted center of mass~\cite{lin2023noise,tang2025collective}. Finally, model V describes active disks with rocking motion~\cite{ruffer1985high,racey1981quasi,guasto2010oscillatory}. Note, that all models except the first feature particle-particle interactions that are purely repulsive and based on volume exclusion. We investigated the different models in a dilute regime, in which no apparent collective behavior takes place, as well as in a dense regime, in which we observe collective behavior.

We have two main findings. First, in the dilute regime, the particle's velocity projected onto its orientation, the so-called effective velocity~\cite{bialke2013microscopic,stenhammar2013continuum,stenhammar2014phase}, scales approximately inversely with the packing fraction~\cite{schnitzer1993theory,cates2015motility,stenhammar2016light,frangipane2018dynamic,arlt2019dynamics}, independently of the assumed microscopic model. Second, in the dense regime, macroscopic phenomena occur that depend on the specific model, namely motility-induced phase separation (MIPS)~\cite{fily2012athermal,buttinoni2013dynamical,cates2015motility,digregorio2018full,van2019interrupted,grossmann2020particle,caporusso2020motility}, travelling waves and global polar order~\cite{vicsek1995novel,chate2008collective,huber2018emergence,chate2020dry,knevzevic2022collective,das2024flocking}, and locally and globally rotating clusters.

\section{Model equations and numerical methods}
In the following, we consider an external field inside a square box. It is consumed by active particles and replenished by diffusion from the boundary according to~\cite{fragkopoulos2021self}
\begin{align}
	\partial_t c(\bm r, t) = D_\mathrm{c}\Delta c(\bm r, t) - k_\mathrm{r}\rho(\bm r, t)c(\bm r, t)\,,
\end{align}
where $D_\mathrm{c}$ is the diffusion constant, $k_\mathrm{r}$ the reaction constant, and $\rho(\bm r, t)$ the local particle density. The external field determines the self-propulsion speed of each particle, whose position and orientation evolve in their general form according to
\begin{align}
\label{eq:langevin-pos}
	\dot{\bm x}_\alpha &= v_\alpha(c)\bm p_\alpha + \frac{1}{\gamma_\mathrm{t}}\bm F_\alpha\\
	\dot{\theta}_\alpha &= \sqrt{2D_\mathrm{r}}\xi_\alpha + \frac{1}{\gamma_\mathrm{r}}T_\alpha\,.
\label{eq:langevin-ori}
\end{align}
Here, $\bm p_\alpha=(\cos\theta_\alpha,\sin\theta_\alpha)^\mathrm{T}$ is the particle's orientation vector, $\gamma_\mathrm{t}$ and $\gamma_\mathrm{r}$ the translational and rotational friction coefficients, respectively, and $D_\mathrm{r}$ the rotational diffusion coefficient. The particles are subject only to rotational but not translational white noise for simplicity, with $\langle\xi_\alpha(t)\rangle=0$ and $\langle \xi_\alpha(t)\xi_\beta(t')\rangle = \delta_{\alpha\beta}\delta(t-t')$. In App.~\ref{app:model-eom}, we list all model equations in more detail together with sketches of the particles and their interactions.

Model I of non-interacting, point-like particles neglects the forces and torques due to particle-particle interactions $\bm F_\alpha$ and $T_\alpha$, respectively. It only contains rotational diffusion as well as self-propulsion depending on the external field~\cite{fragkopoulos2021self} with
\begin{align} 
\label{eq:spp-tanh}
	v_\alpha(c) &= v_\mathrm{min} + \frac{v_\mathrm{max} - v_\mathrm{min}}{2}\bigg[1+\tanh\bigg(\frac{c(\bm x_\alpha, t)-c_\mathrm{typ}}{w}\bigg)\bigg]\,,
\end{align}
where $v_\mathrm{min}$ and $v_\mathrm{max}$ are the minimal and maximal self-propulsion speed and $c_\mathrm{typ}$ and $w$ define the transition between the former two, see Fig.~\ref{fig:spp-vel-vs-conc} in App.~\ref{app:model-eom}.

We extend the description above to model II of sterically repulsive disk-shaped particles via a soft potential of the form
\begin{align}
	U_{\alpha\beta} = \frac{k}{2}(r_\mathrm{cut}-x_{\alpha\beta})^2\,\text{ for } x_{\alpha\beta}<r_\mathrm{cut}\,,
\end{align}
where $k$ controls the repulsion strength, $r_\mathrm{cut}$ is the effective range of the interaction potential, and $x_{\alpha\beta}=|\bm x_\alpha - \bm x_\beta|$ is the distance between two particles. The resulting force is then $\bm F_\alpha=\sum_{\beta\ne\alpha}(-\nabla_\alpha U_{\alpha\beta})$~\cite{fily2012athermal}.

For model III of active dumbbells, we follow~\cite{ostapenko2018curvature,cammann2021emergent} and overlap two beads with radii $a_1, a_2$ and centers a distance $l$ apart. The respective potential energy between two dumbbells then depends on the distances between the individual beads $x_{\alpha\beta}^{(mn)}=|\bm x_\alpha^{(m)}-\bm x_\beta^{(n)}|$, with $m,n\in\left\{1,2\right\}$. We describe the motion of the two beads as a shift of and rotation around the dumbbells midpoint $\bm x_\alpha = \bm x_\alpha^{(1/2)} \mp l\bm p_\alpha/2$. The forces acting on the first and second bead $\bm F_\alpha^{(1/2)}$ effectively act on the levers $\bm x_\alpha^{(1/2)} - \bm x_\alpha=\pm l\bm p_\alpha /2$ and induce the torques $T_\alpha^{(1/2)}=\left\{\pm l\bm p_\alpha/2\times\bm F_\alpha^{(1/2)}\right\}_z$.

For the model IV, we move the center of mass away from the disk's center, similarly to what has been done recently in~\cite{lin2023noise,tang2025collective}. Analogously to model III, we describe the motion of the disk as a shift of and rotation around the center of mass $\bm x_\alpha = \bm x_{\alpha}^\mathrm{gc} + d_\mathrm{gc}(\bm p_\alpha\cos\varphi_\mathrm{gc}+\bm p_\alpha^\perp\sin\varphi_\mathrm{gc})$, where the index ``gc'' denotes the disk's geometric center. Here, $d_\mathrm{gc}$ is the distance between center of mass and geometric center and $\varphi_\mathrm{gc}$ is the offset angle measured from the orientation vector. We consider three types of particles with different offset angles, namely (a) front-heavy disks with $\varphi_\mathrm{gc}=0$, (b) side-back-heavy disks with $\varphi_\mathrm{gc}=3\pi/4$, and (c) back-heavy disks with $\varphi_\mathrm{gc}=\pi$. While the self-propulsion force directly acts on the center of mass, we compute the repulsion force as a function of the geometric centers $\bm F_\alpha^\mathrm{gc}=k\sum_\beta(r_\mathrm{cut}-x_{\alpha\beta}^\mathrm{gc})\hat{\bm x}_{\alpha\beta}^\mathrm{gc}$ and the resulting torque becomes
\begin{align}
	T_\alpha^\mathrm{gc} = \left\{(\boldsymbol x_{\alpha}^\mathrm{gc} - \boldsymbol x_\alpha)\times\boldsymbol F_\alpha^\mathrm{gc}\right\}_z\,.
\end{align}
Note, that the torque would align the particle's distance vector $\boldsymbol x_{\alpha}^\mathrm{gc} - \boldsymbol x_\alpha$ with the repulsion force $\boldsymbol F_\alpha^\mathrm{gc}$, if the interaction was long enough. For (a) and (c), the distance vector is parallel to the particle's orientation vector $\boldsymbol p_\alpha$. In those cases, the resulting reorientation has been referred to in literature as a mechanism for self-alignment, see the appendix A2 in~\cite{baconnier2025self}. 

Finally, for the model V we extend again model II by changing the self-propulsion mechanism to an experimentally motivated rocking motion of the particles~\citep{racey1981quasi,ruffer1985high,guasto2010oscillatory,fragkopoulos2021self}.  During rocking motion, the particles periodically move forward and backward with an offset toward the forward motion. This is e.g. a consequence of the non-reciprocal breast stroke of Chlamydomonas reinhardtii during which the limb-like filaments have to ``stretch forward" at some point in order to perform a new stroke. In our modelling, the new self-propulsion speed is then
\begin{align}
	v^\mathrm{rm}_\alpha(c) = v^\mathrm{off}_\alpha(c) + v^\mathrm{amp}_\alpha(c)\sin(2\pi ft + \varphi_\alpha)\,,
\end{align}
where $v^\mathrm{off}_\alpha(c)$ is the offset speed that follows Eq.~\eqref{eq:spp-tanh}, $v^\mathrm{amp}_\alpha(c)$ is the oscillation's amplitude that is determined from the condition $v^\mathrm{amp}_\alpha(c)/v^\mathrm{off}_\alpha(c)=\mathrm{const.}$, $f$ is the oscillation's frequency, and $\varphi_\alpha$ a constant, but random phase shift.

We performed computer simulations of the models in a square box of size $A_\mathrm{box}=L^2$ with packing fraction $\phi_0=N_\mathrm{p}A_\mathrm{p}/A_\mathrm{box}$. In the following, we use the packing fraction instead of the particle number density $\rho_0=N_\mathrm{p}/A_\mathrm{box}$, because we compare particles of different sizes $A_\mathrm{p}$, namely disks and dumbbells. We used an Euler time-stepping and finite differences for the external field, which is a standard algorithm for solving the heat equation and which is known as forward time-centered space (FTCS), see e.g.~\cite{anderson2020computational}. For the particles, we used an Euler-Maruyama time-stepping scheme. We use a combination of a Verlet and cell neighbor list to compute particle-particle interactions efficiently~\cite{verlet1967computer,allen2017computer}. The neighbor list subdivides the system into cells the size of one particle plus a buffer zone. To determine possible neighbors of a particle, only particles in neighboring cells need to be considered. The buffer zone reduces the need to update the neighbor list to less then once a time step. The simulation code as well as the numerical methods are largely inspired by a tutorial presented at the KITP ACTIVE20 workshop~\cite{sknepnek2020tutorial}. Motivated by the experiment~\cite{fragkopoulos2021self}~that shows a radially symmetric geometry, we chose Dirichlet boundary conditions ($c=c_\mathrm{sat}$) for all grid points of the external field outside a circle, see Fig.~\ref{fig:models-sketch}. On the other hand, we implemented periodic boundary conditions for the particles to avoid aggregation at the boundary. This means that particles can move also through areas in which the external field is kept constant. For the statistics involving the external field, we explicitly ignored particles that were outside the circle during measurements. The time evolution of the external field additionally requires the local particle density. For that, we used a simple step function that assigns particles to the nearest grid point. Vice versa, we used bilinear interpolation between grid points to compute the self-propulsion velocity from the external field. For more details on parameter values and observables see App.~\ref{app:parameters} and App.~\ref{app:observables}, respectively.

\begin{figure}[!htb]
\includegraphics[width=\columnwidth]{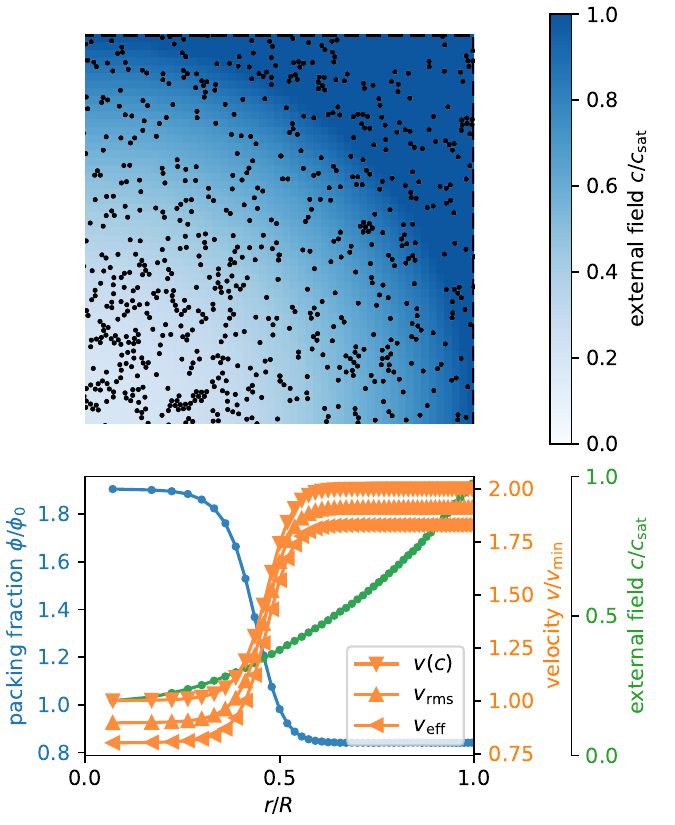}
\caption{Snapshot of one quadrant and radial profiles of the system of sterically repulsive active disks (model II, $\phi_0=0.1$). The particles consume the external field which leads to a gradient that decreases towards the center ($r=0$). There, the particles slow down and aggregate. The particles' root-mean square velocity $v_\mathrm{rms}$ is smaller than the self-propulsion speed $v(c)$ but still larger than the effective velocity $v_\mathrm{eff}$.}
\label{fig:snapshot-and-radial-profiles}
\end{figure}

\section{Dilute regime with external field}
In the dilute regime, the system exhibits very similar dynamics for all five models, see example snapshot of model II in Fig.~\ref{fig:snapshot-and-radial-profiles}. 
We measure the radial profiles of the external field $c(r)$, the packing fraction $\phi(r)=n_\mathrm{p,shell}(r)A_\mathrm{p}/A_\mathrm{shell}$, and the particle velocity $v(r)$ in shells of same area for better statistics in the center. In the steady state, we observe a radial gradient of the external field. In the center, where the external field is low, the particles move more slowly due to their velocity dependence~$v_\alpha(c)$~on the external field, see Eq.~\ref{eq:spp-tanh}. This in turn causes the particles to spend more time in the center increasing the particle density there. In all models that feature steric interactions, and thus volume exclusion, the particles have a root mean square velocity $v_\mathrm{rms}=\sqrt{\langle \bm v_\alpha^2\rangle}$ that is smaller than their self-propulsion speed $v_\alpha(c)$ due to the interactions. Similarly, the particle velocity projected onto their orientation, called the effective velocity $v_\mathrm{eff}=\langle \bm v_\alpha\cdot\bm p_\alpha\rangle$~\citep{bialke2013microscopic}, is even smaller than the root mean square velocity because it ignores the motion orthogonal to the particle's orientation.

\begin{figure}[!htb]
\includegraphics[width=\columnwidth]{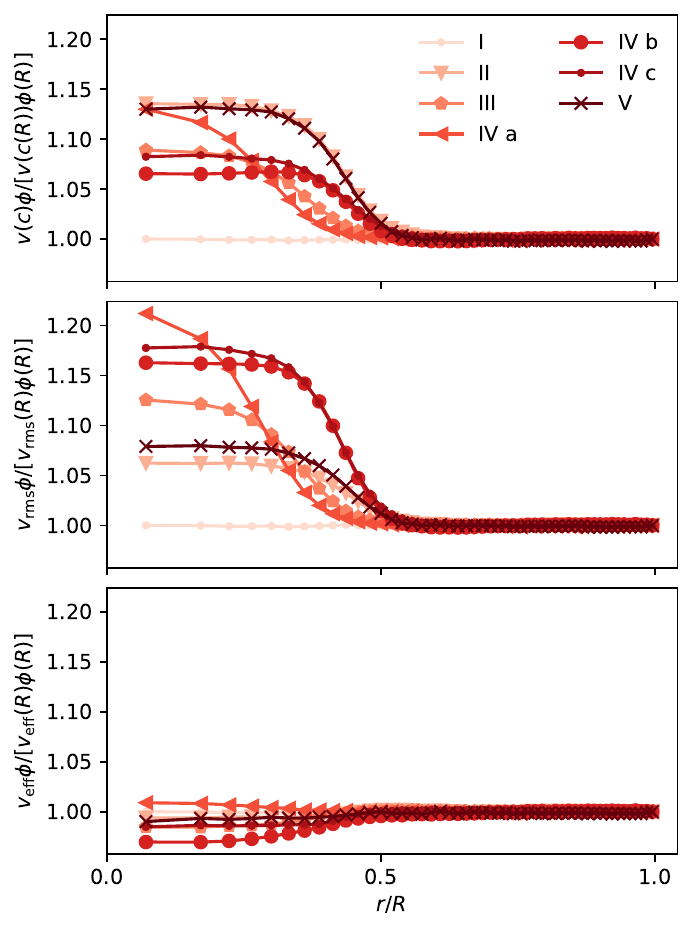}
\caption{The product of velocity and packing fraction shows the deviation from inverse scaling behavior for the self-propulsion speed $v(c(r))$ and root mean square velocity $v_\mathrm{rms}(r)$. In comparison, the effective velocity $v_\mathrm{eff}(r)$ is close to the inverse scaling behavior. The three curves for model IV ($d_\mathrm{gc}=0.4$) correspond to a shifted center of mass to the front (a: $\varphi_\mathrm{gc}=0$), diagonal back (b: $\varphi_\mathrm{gc}=3\pi/4$), and back (c: $\varphi_\mathrm{gc}=\pi$). We used the packing fraction $\phi_0=0.1$ for all models except model IV a for which we used $\phi_0=0.08$ to avoid clustering.}
\label{fig:scaling}
\end{figure}

The radial profiles in Fig.~\ref{fig:snapshot-and-radial-profiles} already indicate that packing fraction and particle velocity scale roughly inversely with each other, i.e. $\phi(r)\sim 1/v(r)$. This hypothesis is true for non-interacting active particles in an external field, see~\cite{schnitzer1993theory,cates2015motility,stenhammar2016light,frangipane2018dynamic,arlt2019dynamics}. 
It makes sense intuitively, in that particles spend more time in regions of low velocity than regions of high velocity. Consequently, regions of low velocity should have a higher packing fraction. One finds this relation also analytically with the help of the Fokker-Planck equation, see App.~\ref{app:steady-state-inverse-scaling}.

However, it becomes more intricate in the case of steric particle interactions. When two particles collide, their motion is changed in a twofold manner. The particles slow down along their direction of self-propulsion and they push each other to the side, resulting in an additional motion perpendicular to their direction of self-propulsion~\citep{bialke2013microscopic}. In this case the self-propulsion velocity $v_\alpha(c(r))$ and the root mean square velocity $v_\mathrm{rms}(r)$ deviate systematically from the inverse scaling behavior, see Fig.~\ref{fig:scaling}. The effective velocity $v_\mathrm{eff}(r)$ on the other hand displays a reasonable approximation to the inverse scaling behavior for all models. Note, that the deviations from the inverse scaling behavior decrease the more dilute the system is, see Fig.~\ref{fig:scaling-model-IV-b} in the appendix.

\begin{figure}[!htb]
\includegraphics[width=\columnwidth]{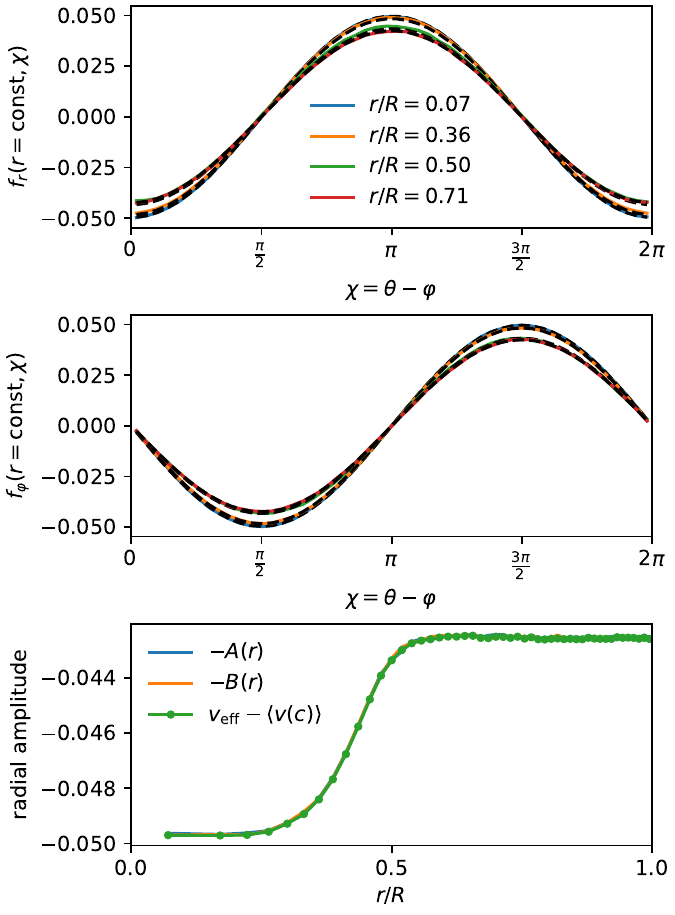}
\caption{Radial and azimuthal components of the conditional force average $f_r(r=\text{const},\chi)$ and $f_\varphi(r=\text{const},\chi)$ are cosine and sine functions of the angular difference $\chi=\theta-\varphi$. They vary in amplitude depending on the radial distance from the center of the box. The amplitudes $A(r)$ and $B(r)$ are equal and agree with the deviation of the effective velocity from the self-propulsion speed $v_\mathrm{eff}-v(c)$. The data shown is from simulations of active disks (model II, $\phi_0=0.1$) as in Fig.~\ref{fig:snapshot-and-radial-profiles}.}
\label{fig:force-fields}
\end{figure}

To recover the inverse scaling relation analytically, one needs to take into account the interaction forces. Here, we follow a similar argument as in ~\cite{bialke2013microscopic,van2019interparticle} and consider the one-particle probability density function $\psi(\bm x, \theta, t)=\langle\delta(\bm x_\alpha - \bm x)\delta(\theta_\alpha - \theta) \rangle$. It evolves in time according to the Fokker-Planck equation, see App.~\ref{app:derivation-FPE}
\begin{align}
	\partial_t \psi(\bm x, \theta, t) = &-\nabla_{\bm x}\cdot\left[\left(v(c)\bm p + \frac{1}{\gamma_\mathrm{t}} \bm f(\bm x, \theta, t)\right)\psi(\bm x, \theta, t)\right] \nonumber\\
	& - \partial_\theta \left[\left( \frac{1}{\gamma_\mathrm{r}} \tau(\bm x,\theta,t) - D_\mathrm{r}\partial_\theta\right)\psi(\bm x, \theta, t)\right]\,,
\label{eq:fokker-planck}
\end{align}
where $\bm f(\bm x, \theta, t)=\langle\bm F_\alpha |\bm x_\alpha=\bm x, \theta_\alpha=\theta\rangle$ and $\tau(\bm x, \theta, t)=\langle T_\alpha |\bm x_\alpha=\bm x, \theta_\alpha=\theta\rangle$, 
for which $\langle\cdot|\bm x_\alpha=\bm x,\theta_\alpha=\theta\rangle$ is an unclosed conditional average that depends on the conditional two-particle probability density function $\psi(\bm x', \theta', t|\bm x, \theta)$. Instead of attempting a physics-motivated closure as was done in~\cite{bialke2013microscopic,van2019interparticle}, we directly measure the conditional averages, see Fig.~\ref{fig:force-fields}. Because of radial symmetry, we expect them to depend only on the angular difference $\chi=\theta-\varphi$ of the particle orientation $\theta$ and the azimuthal angle of the particle position $\varphi$. We find that the radial and azimuthal components of the conditional force average are cosine and sine functions with amplitudes $A(r)$ and $B(r)$ that depend on the radius:
\begin{align}
	f_r(r,\chi,t)&=\bm f(\bm x, \theta)\cdot\bm e_r\nonumber\\
	&= -A(r)\cos\chi\\
	f_\varphi(r,\chi,t)&=\bm f(\bm x, \theta)\cdot\bm e_\varphi\nonumber\\
	&= -B(r)\sin\chi\,.
\end{align}
Actually, both amplitudes are approximately the same, i.e. $A(r)=B(r)$, which results in a force that acts antiparallel to the particle orientation:
\begin{align}
	\bm f(\bm x, \theta) &= -A(r)\cos\chi\bm e_r - B(r)\sin\chi \bm e_\varphi \nonumber\\
	&=-A(r)\bm p\,.
\end{align} 
This means that particle interaction forces on average slow down the particle self-propulsion, resulting in the effective velocity described above. Interestingly, our numerical closure of the Fokker-Planck equation does not feature an additional contribution to a diffusive motion as has been proposed in~\citep{bialke2013microscopic}. In App.~\ref{app:eff-vel-diff} we show that this discrepancy occurs because any positive contribution to a diffusive motion due to steric interactions is numerically small in contrast to either rotational diffusion or the slowdown due to the same interactions. 

\begin{figure}[!htb]
\includegraphics[width=\columnwidth]{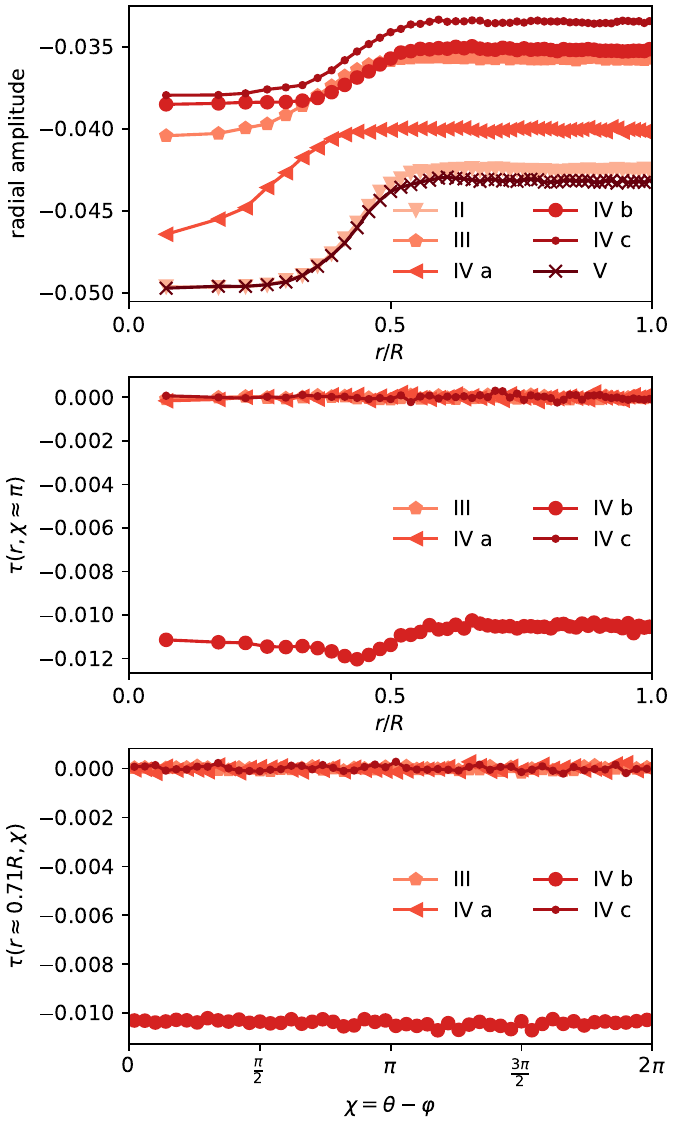}
\caption{Radial amplitude $A(r)$ of the conditional force average as well as the conditional torque average $\tau(r,\chi)$ for different models and same parameters as in Fig.~\ref{fig:scaling}.}
\label{fig:force-torque-contribution}
\end{figure}

\begin{figure*}[!htb]
\includegraphics[width=\textwidth]{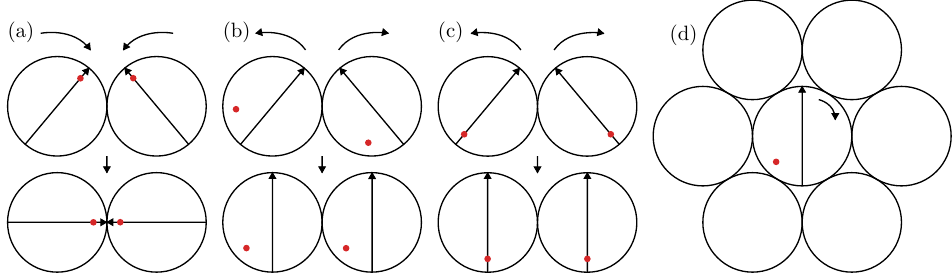}
\caption{Pairwise interactions of disks with shifted centers of mass (model IV*) lead to (a) trapped or (b) and (c) aligning particles, see also movies in SM~\cite{supplementalMaterial}. When particles form a cluster, (d) the hexagonal ordering emulates closed boundaries: the six contact points approximate a circle. For disks with shifted centers of mass to the side (model IV b*) this leads to a clockwise rotation. This is in line with our observations of overall rotating clusters, see Fig.~\ref{fig:clash-of-clusters-characterization}.}
\label{fig:pairwise-interaction}
\end{figure*}

Comparing the radial amplitude $A(r)$ for different models, see Fig.~\ref{fig:force-torque-contribution}, we find that it has the same shape but different magnitude for almost all models. For the disks with center of mass shifted to the front (model IV a) the radial amplitude decreases more steeply towards the center. This enhanced effective reduction of the particle speed is probably due to the type of interaction in which particles turn towards each other, see Fig.~\ref{fig:pairwise-interaction}. Considering the conditional torque average $\tau(\bm x, \theta, t)$ in models III and IV a, b, and c, we find it to be zero in all but one case, see Fig.~\ref{fig:force-torque-contribution}. For disks with shifted center of mass to the side and back (model IV b), the conditional torque average is constant and non-zero as a function of the angular difference $\chi$ and close to constant with respect to the radial distance to the center. Consequently, the contribution vanishes in Eq.~\eqref{eq:fokker-planck} when the derivative with respect to the particle orientation is taken.

In summary, all studied models show an effective velocity that is close to scaling inversely with the packing fraction. While it does not explain the experimentally found scaling behavior of $v\propto\rho^{-0.5}$~\citep{fragkopoulos2021self}, it is an extension to earlier studies~\cite{bialke2013microscopic,stenhammar2016light,
frangipane2018dynamic,arlt2019dynamics}.

\begin{figure*}[!htb]
\includegraphics[width=\textwidth]{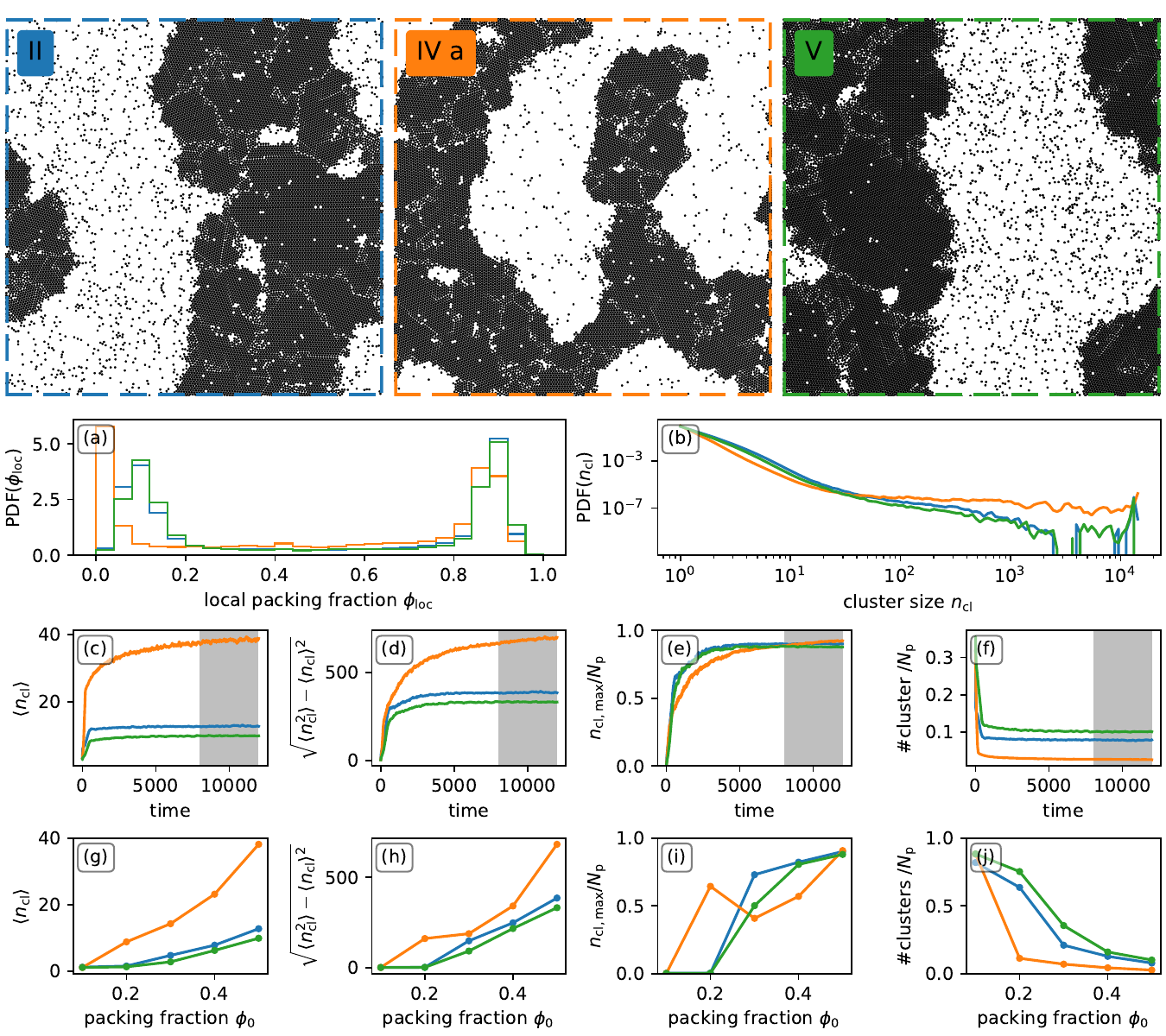}
\caption{First row: Snapshots of disks (model II*), front-heavy disks (model IV a*), and disks with rocking motion (model V*) display motility-induced phase separation for high packing fraction ($\phi_0=0.5, v_0=0.5$). The star at the model number indicates that the external field was omitted. Second row: The front-heavy disks turn toward each other and allow for less particles in the dilute phase than the other two models. This can be seen in the left peak of the (a) distribution of the local packing fraction $\phi_\mathrm{loc}$ and the approximately constant (b) likelihood of large cluster sizes $n_\mathrm{cl}$. Third row: The time evolution of the (c) mean cluster size $\langle n_\mathrm{cl}\rangle$, the (d) standard deviation $\sqrt{\langle n_\mathrm{cl}^2\rangle-\langle n_\mathrm{cl}\rangle^2}$, the (e) cluster size of the largest cluster $n_\mathrm{cl,max}$ and the (f) number of clusters show that the front-heavy disks form more stable clusters that take longer to equilibrate. The shaded grey area marks the time average for all other quantities displayed in this plot. Fourth row: Varying (g-j) packing fraction further reveals that front-heavy disks form clusters already for lower packing fractions than the other two models.}
\label{fig:mips-characterization}
\end{figure*}

\section{Dense regime without external field}

In contrast to the dilute systems above, the dense systems can show very different collective behavior depending on the details of the model. These collective behaviors include motility-induced phase separation (MIPS)~\cite{fily2012athermal,buttinoni2013dynamical,cates2015motility,digregorio2018full,van2019interrupted,grossmann2020particle,caporusso2020motility},  travelling waves and global polar order~\cite{vicsek1995novel,chate2008collective,huber2018emergence,chate2020dry,knevzevic2022collective,das2024flocking}, as well as locally and globally rotating clusters. In the following, we first study the systems without external field but with constant self-propulsion speed and add the external field later to compare how it influences the collective behavior. We asterisk the model names, whenever we mean the system without external field and with constant self-propulsion speed or constant offset velocity $v_\alpha(c)=v^\mathrm{off}_\alpha(c)=\mathrm{const.}=v_0$.

\subsection{Motility-induced phase separation}
We observe motility-induced phase separation for disks (model II*), front-heavy disks (model IV a*), and disks with rocking motion (model V*), see Fig.~\ref{fig:mips-characterization}. MIPS is a collective phenomenon during which active particles separate into regions of high and low density. For active particles, a region of high density can form even when the particles interact purely repulsively. During a collision, particles slow down which can trigger a feedback-loop of other particles joining the collision and also slowing down until a large cluster-like structure forms~\cite{fily2012athermal,buttinoni2013dynamical,cates2015motility,digregorio2018full,van2019interrupted,grossmann2020particle,caporusso2020motility}  . The formation of a cluster may be, however, suppressed if additional effects such as alignment of the particle orientation takes place, as we explain below.

Model II* is essentially the paradigm model for studying MIPS and has been discussed in detail in the literature mentioned above. Hence in the following, we focus on comparing the size and stability of the clusters of the three different models in which we observe MIPS, i.e. models II*, IV a*, and V*. 

Note that all three models have in common that the particles' shape is circular which results in dense hexagonal packing. In contrast to the other two models, the front-heavy disks from model IV a* are additionally subject to a torque that turns particles towards each other, see Fig.~\ref{fig:pairwise-interaction}(a) and movies in the Supplemental Material~\cite{supplementalMaterial}. Finally, note that particles from model V* perform a rocking motion, i.e. they move forward and backward periodically with a bias toward the forward motion.

Our first observation is that front-heavy disks that undergo MIPS form regions of low packing fraction that have less particles than corresponding regions of disks with and without rocking motion, see Fig.~\ref{fig:mips-characterization}. This becomes apparent in snapshots of the full system as well as in the local packing fraction distribution $\mathrm{PDF}(\phi_\mathrm{loc})$, see Fig.~\ref{fig:mips-characterization}(a) and App.~\ref{app:observables} for the definition. The local packing fraction distribution takes the characteristic bimodal shape~\cite{digregorio2018full,grossmann2020particle}~for all three models, but notably shifts the low packing fraction peak towards zero for the front-heavy disks. In addition, the mean cluster size $\langle n_\mathrm{cl} \rangle$~\cite{ginelli2010large,fily2012athermal,buttinoni2013dynamical,weitz2015self,van2019interrupted,caporusso2020motility}  and the total number of clusters (Fig.~\ref{fig:mips-characterization}(c) and (f)) is significantly higher and lower, respectively, which reflects the amount of single-particle-clusters in the system. An explanation for this observation is the two-particle interaction behavior of front-heavy disks mentioned above. Front-heavy disks at the border of the dense phase tend to turn towards the inside of the cluster and need more time to rotate away due to rotational diffusion. 

This explanation already hints at our second observation that front-heavy disks form more stable clusters than disks with and without rocking motion. The time evolution of the mean and standard deviation of the cluster size $\langle n_\mathrm{cl} \rangle$ and $\sqrt{\langle n_\mathrm{cl}^2 \rangle-\langle n_\mathrm{cl} \rangle^2}$ as well as of the size of the largest cluster $n_\mathrm{cl,max}$, see Fig.~\ref{fig:mips-characterization}(c-e), show that models II* and V* reach the statistical steady state after about a third of the displayed simulation time, while model IV a* does not quite reach the steady state at all. Note that despite no steady state is reached, we took a time average for the statistics over the last third of the simulation time marked by the shaded grey area for practical reasons. Furthermore, the cluster size distribution $\text{PDF}(n_\mathrm{cl})$ (Fig.~\ref{fig:mips-characterization}(b)) is approximately constant for sizes in between $10^2$ to $10^4$ for front-heavy disks while it, except for a peak at $10^4$, decays and partially vanishes in that range for disks with and without rocking motion. This means that large cluster sizes can persist over long times for the former and merge more quickly to a few very large clusters for the latter. Last but not least, the parameter scan over the packing fraction reveals that clustering occurs already for smaller global packing fractions for front-heavy disks than for disks with and without rocking motion (Fig.~\ref{fig:mips-characterization}(g-j)). The percentage of particles inside the largest cluster $n_\mathrm{cl,max}/N_\mathrm{p}$ surprisingly is larger for a packing fraction of $\phi_0=0.2$ than for $\phi_0=0.3$ for front-heavy disks. This may be because the system with lower packing fraction starts with less nucleation cores which take less time to merge into a large cluster, see movies in the SM~\cite{supplementalMaterial}.

We find only small differences between disks with and without rocking motion, namely smaller cluster size statistics and a larger number of clusters for the latter. The small deviation may come from the rocking motion that allows particles to back out of a parking space.

\begin{figure*}[!htb]
\includegraphics[width=\textwidth]{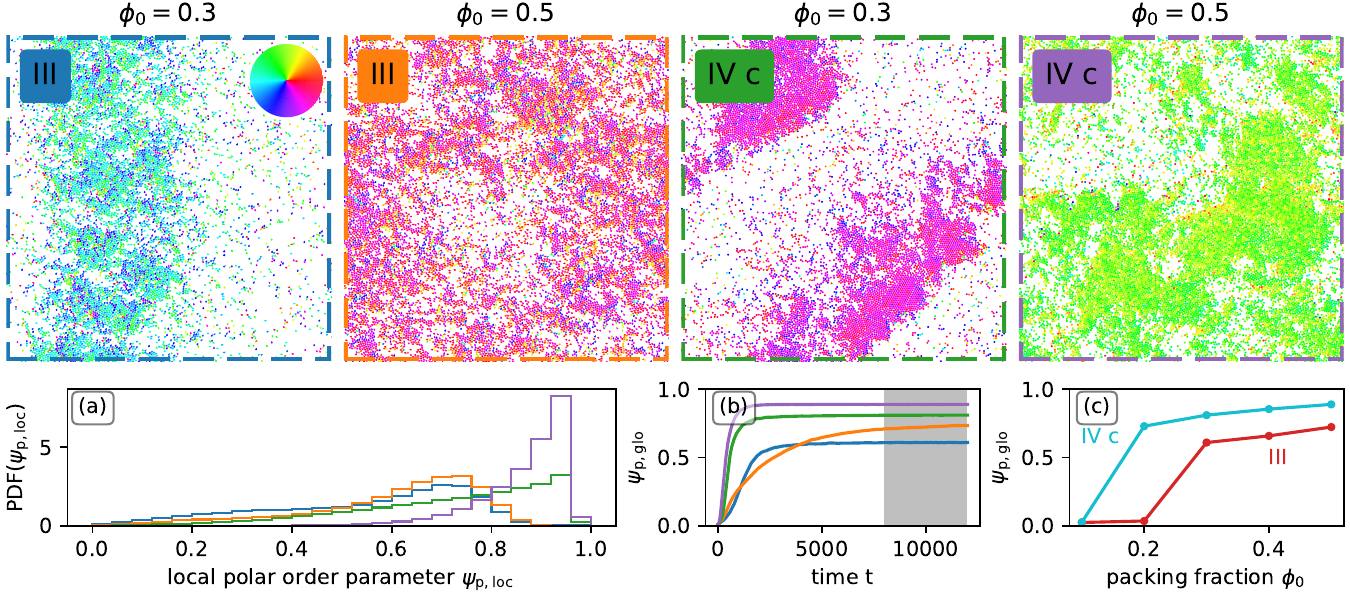}
\caption{First row: Snapshots of dumbbells (model III*) and back-heavy disks (model IV c*) display the formation of travelling waves for medium and global ordering for high packing fractions ($\phi_0=0.3, 0.5,v_0=0.5$, the color code shows the particle orientation angle). Second row: The (a) local polar order parameter $\psi_\mathrm{p,loc}$ is slightly larger for back-heavy disks than for dumbbells and much more pronounced for the global order case. The (b) time evolution of the global polar order parameter $\psi_\mathrm{p,glo}$ reveals a longer equilibration time for the global order state of dumbbells than for the other cases. Moreover, back-heavy disks show (c) a stronger ordering than dumbbells for all studied packing fractions.}
\label{fig:bands-characterization}
\end{figure*}

\begin{figure*}[!htb]
\includegraphics[width=\textwidth]{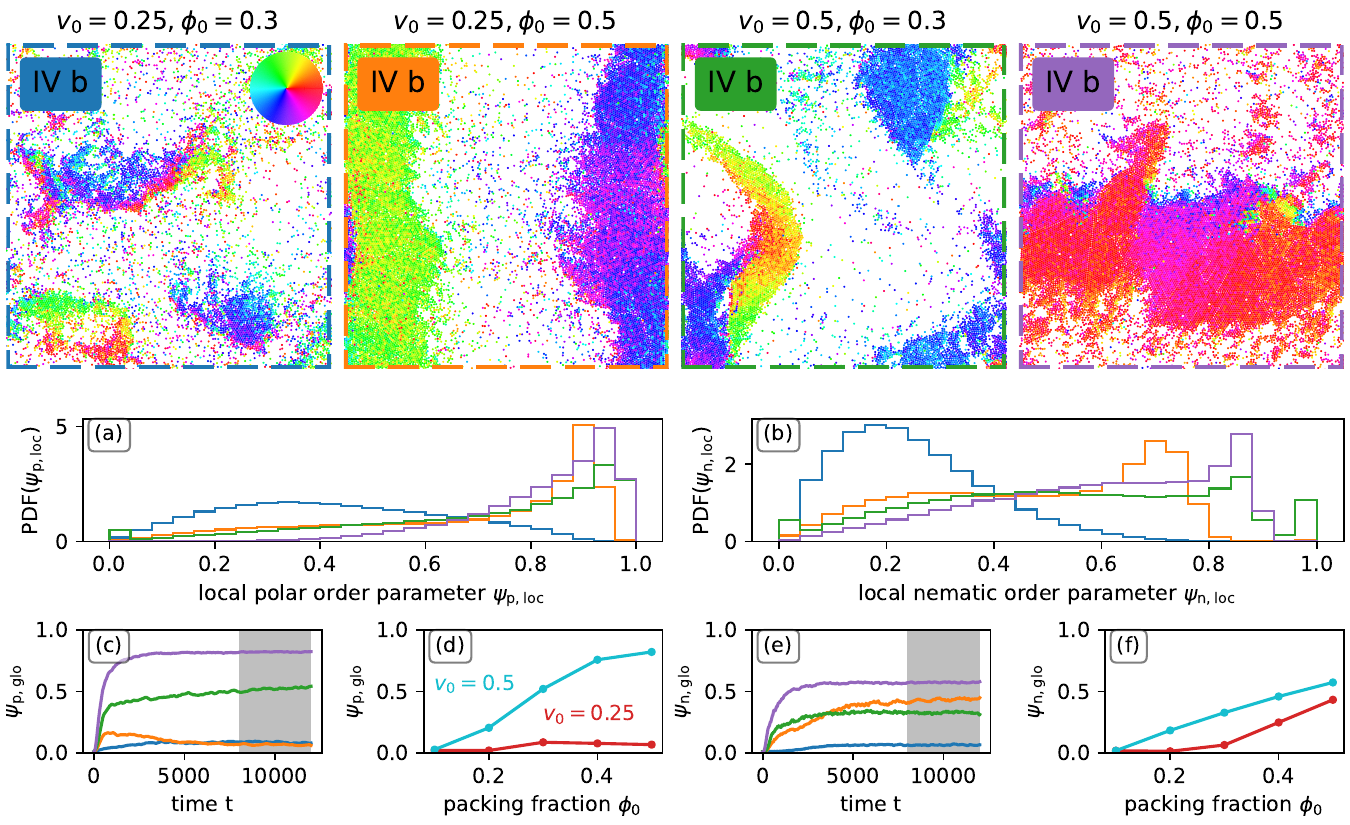}
\caption{First row: Snapshots of slow (blue and orange) and fast (green and purple) side-back-heavy disks (model IV b*) display the different states of ``clashing clusters". From left ro right: small clusters rotating around centers outside the cluster and dispersing due to collisions, two oppositely oriented bands that bounce off each other while the mean orientation rotates periodically, two smaller clusters that rotate around centers outside the cluster in opposing direction, and one large cluster that fills the whole domain and rotates. Second row: The distribution functions of the (a) local polar and (b) nematic order parameters $\text{PDF}(\psi_\mathrm{p,loc})$ and $\text{PDF}(\psi_\mathrm{n,loc})$ show a high polar order in all but the first case. Third row: High (c-d) global polar order ($\psi_\mathrm{p,glo}$) only occurs for high velocity, while high (e-f) global nematic ($\psi_\mathrm{p,glo}$) order also occurs for low velocity, which indicates the formation of bands with opposite polar order.}
\label{fig:clash-of-clusters-characterization}
\end{figure*}

\begin{figure*}[!htb]
\includegraphics[width=\textwidth]{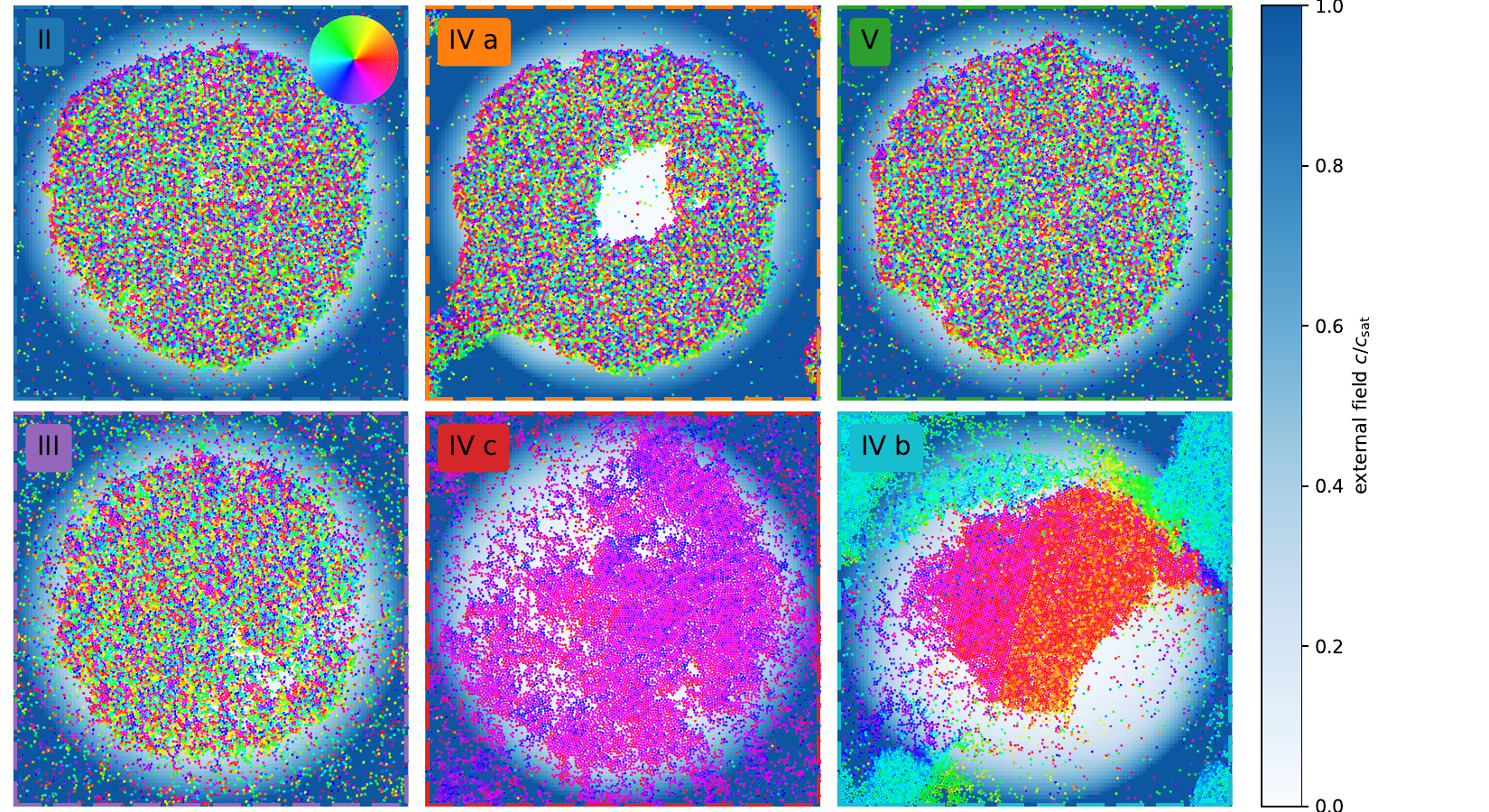}
\caption{First row: Snapshots of disks (model II), front-heavy disks (model IV a), and disks with rocking motion (model V) inside an external field with radial symmetry display MIPS. The clusters are centered and also radially symmetric. Front-heavy disks need longer times to reach the statistically steady state with a fully filled cluster and can sustain stable holes. Second row: Snapshots of dumbbells (model III), back-heavy disks (model IV c) and side-back-heavy disks (model IV b) inside an external field. The dumbbells form a cluster similar to MIPS without the stable hexagonal packing and constantly fluctuate. The back-heavy disks display global polar order with a smaller moving and hence denser core in the center. The side-back-heavy disks form two clusters, inside and outside the center, that rotate around a center and clash into each other. All simulations have the packing fraction $\phi_0=0.5$.
}
\label{fig:collective-behavior-with-external-field}
\end{figure*}

\subsection{Travelling waves and global polar order}
We observe travelling waves and global polar order for dumbbells (model III*) and back-heavy disks (model IV c*), see Fig.~\ref{fig:bands-characterization}. Travelling waves are again a collective phenomenon during which active particles separate into regions of low and high packing fraction. In comparison to MIPS, however, the regions of high packing fraction are elongated clusters that span through the whole system and move in the direction perpendicular to their extension. The active particles in the regions of high packing fraction on average point in the direction of the moving cluster. They have been observed for the Vicsek model~\cite{vicsek1995novel,chate2008collective,chate2020dry}, active particles with non-reciprocal orientational interactions~\cite{knevzevic2022collective,das2024flocking}, and in experiments~\cite{huber2018emergence}. Travelling waves differ from bands or lanes that are elongated clusters with particles moving in the direction parallel to the extension. These bands or lanes are e.g. observed for self-propelled rods~\cite{ginelli2010large,weitz2015self,grossmann2020particle}. On the other hand, global polar order is a collective phenomenon during which almost all active particles move uniformly in one direction. In this case, no clear structure of a cluster can be identified. 

Note that the dumbbells from model III* consist of overlapping beads of which one is larger than the other, whereas the back-heavy particles from model IV c* have a circular shape. Particles of both models have a center of mass that is shifted with respect to the geometric center of the (larger) disk. For better comparison, we shifted the center of mass in the same manner for the disks as for the larger bead of the dumbbells ($\varphi_\mathrm{gc}=\pi, d_\mathrm{gc}=0.25$) so that only a difference in shape persists. The shift in the center of mass this time leads to a parallel alignment of the particle orientations, see Fig.~\ref{fig:pairwise-interaction}(c) and movies in the SM~\cite{supplementalMaterial}.

Our first observation is that the circular shape of the disks facilitates the polar ordering of the particles in comparison to the anisotropic shape of the dumbbells. This becomes apparent in the distribution function of the local polar order parameter $\text{PDF}(\psi_\mathrm{p,loc})$ as well as the time evolution and global packing fraction dependence of the global polar order parameter $\psi_\mathrm{p,glo}$, see Fig.~\ref{fig:bands-characterization}(a-b) and App.~\ref{app:observables} for a definition. The distribution function shows a larger local polar order parameter for back-heavy disks than dumbbells for both phenomena. Furthermore, the time evolution of the global polar order parameter demonstrates that it takes dumbbells a longer time to reach a statistically steady state. Finally, the system of front-heavy disks enters an ordered state for lower global packing fractions. 

One reason for these observations may be the ability of disks to enter a dense hexagonal packing, which is more difficult for the dumbbells with their anisotropic shape. A dense packing of disks leads to close and repeated contact between the disks aligning them more strongly while keeping them in place. The dumbbells are actually able to enter a similar but transitive state during which the dumbbells form a dense cluster-like structure, see the movies in the SM~\cite{supplementalMaterial} and Fig~\ref{fig:collective-behavior-with-external-field} for comparison. This cluster-like structure fluctuates strongly as the dumbbells penetrate the structure quite easily. The dumbbells' motion through the cluster constantly reorients them in whichever direction leads to a small gap. This motion actually reduces their alignment, which is contrary to the enhancement of alignment for disks. The movie shows that the dumbbells outside this structure start to align and ``wash against" the cluster, tearing at its boundary until the cluster shrinks enough to disperse. This explains why the dumbbells need a longer time to reach a statistical steady state.

Our second observation is that travelling waves are, in terms of the global packing fraction, an intermediate phenomenon that transitions into a state of global polar order when increasing the packing fraction. That becomes apparent in the snaphots for intermediate and high packing fraction, see Fig.~\ref{fig:bands-characterization}. The travelling waves leave a wake behind the sharp front, which potentially washes out the front in the case of increasing packing fraction. In other words, both the travelling waves and the global polar order may be the same phenomenon only bounded by the system size.

\subsection{Clash of clusters}
We observe locally and globally rotating clusters for side-back-heavy disks (model IV b*), see Fig.~\ref{fig:clash-of-clusters-characterization}.
They are once more a collective phenomenon during which the active particles separate into regions of low and high packing fraction. As before with the travelling waves, the active particles align with each other in the dense regions, see Fig.~\ref{fig:pairwise-interaction}(b) and movie in the SM~\cite{supplementalMaterial}. However, they additionally collectively perform a clockwise rotating motion around a barycenter that is outside the dense region. Qualitatively, this phenomenon can be explained by studying an active particle confined by other particles surrounding it, see Fig.~\ref{fig:pairwise-interaction}(d) and movie in the SM~\cite{supplementalMaterial}. In a dense hexagonal packing, the particle encounters an approximately circular confinement. Forward motion in this confinement leads to a reorientation of the particle, because the particle's center of mass ideally would be positioned where the particle moves against the boundary. However, the forward motion depends on the particle's orientation vector which rotates with the particle. Hence, the particle's rotation persists. Inside the dense region, almost all particles perform the same rotation which leads to an overall rotation of the cluster. Note, that this is different from, e.g., circle swimmers, where a single particle already performs a rotating motion~\cite{liao2018clustering,liao2021emergent}.

Depending on size, the clusters remain relatively stable while they rotate through the system. When two or more clusters collide, they deform and slide off each other. This ``clash of clusters" varies in form depending on the self-propulsion speed and the global packing fraction. We observed statistically steady states of a clash of four or two clusters, a pair of rotating bands, and a global collective rotation of a single cluster, see Fig.~\ref{fig:clash-of-clusters-characterization}.

In all cases but the clash of four clusters ($v_0=0.25, \phi_0=0.3$), there is a strong local polar order, which becomes evident from the distribution function of the local polar order parameter $\text{PDF}(\psi_\mathrm{p,loc})$, see Fig.~\ref{fig:clash-of-clusters-characterization}(a). Therefore, we need more than the local order parameter to distinguish the different cases.

When the velocity is low and the packing fraction high ($v_0=0.25,\phi_0=0.5$), we observe two bands that span the whole system. They rotate with a phase shift  to each other of approximately $\pi$ and collide with each other repeatedly at two separate points, see movies in the SM~\cite{supplementalMaterial}. They have approximately opposite polarity. This can be seen when studying the global polar and nematic order parameters $\psi_\mathrm{p,glo}$ and $\psi_\mathrm{n,glo}$, see Fig.~\ref{fig:clash-of-clusters-characterization}(d,f) and App.~\ref{app:observables} for a definition. While the global polar order parameter remains low for all studied packing fractions, the global nematic order parameter increases towards higher packing fractions. These bands may be the result of multiple factors. One is probably the box size that restricts the size of a single cluster, while the others are the reorientation and self-propulsion speed that restrict the radial distance from the common barycenter. 

\section{Dense regime with external field}
We observe MIPS, global polar order, and a clash of clusters also in the system with external field and radial symmetry, see Fig.~\ref{fig:collective-behavior-with-external-field}. All three differ in their details from the collective behavior observed in the system without external field.

For MIPS in models II, IV a, and V, the clusters nucleate at a radius that is roughly a quarter the box length, see movies in the SM~\cite{supplementalMaterial}. At this radius, the self-propulsion velocity, which depends on the external field, steeply increases outward from low to high velocity. The steep increase facilitates the formation of clusters, because the slowdown is matched by quickly arriving particles. From the nucleation point, the clusters grow towards the center until a circular area is filled with particles. Again, the equilibration time for the front-heavy disks is much longer than for disks with and without rocking motion which is why a hole in the cluster center can persist.

Interestingly, the dumbbells (model III) also form an isotropic cluster instead of aligning with each other. The cluster differs from the cases above in that it constantly fluctuates and does not arrest the particles, see movies in the SM~\cite{supplementalMaterial}. This is probably due to the shape of the dumbbells that doesn't allow a perfect hexagonal packing.

For global polar order in model IV c, the aligned disks move constantly in the same direction. They move more slowly through the center of the box and faster at the edges. This leads to a delay in the center that keeps the packing fraction constantly higher than outside the center. A band probably cannot form, because it would constantly be cut off at the sides of the chamber where the particles moves faster.

Finally, for the clash of clusters in model IV b, we observe an inside and an outside cluster that rotate around the center and the corners of the box, respectively. Both move clockwise but with a phase shift.

\section{Conclusion}
In order to understand to which extent active particle models show (non)universal behavior, we compared the statistics and collective behavior of five different particle models inside an external field. We considered point-like particles, sterically interacting disks, dumbbells, disks with shifted center of mass, and disks with rocking motion. 

For all models except the point-like particles that do not interact at all, the only type of particle-particle interaction was steric repulsion. On the one hand, this similarity between the models led, in the dilute system, to the same inverse scaling behavior of the effective velocity with the packing fraction. To understand the observed behavior, we directly measured the conditional force and torque averages. We found that on average the particles were effectively slowed down in opposing direction of their orientation, almost independent of the details of the interaction. We thereby extended the observation from earlier literature to particles of different shape, center of mass, and self-propulsion mechanism. Furthermore, the similarity in the interactions led, in the dense system, always to a separation in regions of low and high packing fraction. We found that front-heavy disks undergo motility-induced phase separation for lower packing fractions than disks with and without rocking motion and that they form more stable clusters.

On the other hand, the differences between the models in shape and center of mass enabled an additional particle reorientation and, therefore, different collective motion. We found that both dumbbells and back-heavy disks can form travelling waves and global polar order with increasing packing fraction. Without an external field, they differed primarily in the magnitude of the polar ordering. In contrast, dumbbells exhibited a dense but constantly fluctuating cluster when subject to the external field. Furthermore, the side-back-heavy disks formed rotating, stable but deformable clusters.

The results highlight that modeling the collective behavior in experimental systems with the help of active particle models requires rather precise knowledge of the relevant microscopic interactions. 
Future research may extend this analysis across models to long-range interactions such as hydrodynamic interactions as has been done in~\cite{theers2018clustering}.

\begin{acknowledgments}
Calculations were performed using the emil-cluster and the festus-cluster of the 
Bayreuth Centre for High Performance Computing (https://www.bzhpc.uni-bayreuth.de),
funded by the Deutsche Forschungsgemeinschaft (DFG, German Research Foundation) - 422127126 and 523317330, respectively.

We thank Alexandros Fragkopoulos and Oliver Bäumchen for helpful discussions on the experimental system and its analysis. Further, we thank Rastko Sknepnek for insights on and providing the basis of the numerical code. Finally, we thank Lukas Bentkamp, Gabriel B. Apolinário, Kim Kreienkamp, and Henning Reinken for numerous very helpful discussions on the theory and statistical analysis.
\end{acknowledgments}

\newpage

\appendix

\section{Parameters}
\label{app:parameters}
The five models feature many parameters which makes a full parameter study unfeasible. Here, we list the parameters that were used to create the figures in the main text and in the appendix. For the equations of motion see the corresponding subsection.

If not specified otherwise, we used the domain size $L=150$ (domain area $A_\mathrm{b}=L^2$) and set the packing fraction $\phi_0=N_\mathrm{p}\frac{A_\mathrm{p}}{A_\mathrm{b}}$ instead of the particle number $N_\mathrm{p}$ to account for the different particle areas $A_\mathrm{p}$ of the disks and dumbbells. In model I, we also used the particle area $A_\mathrm{p}=\pi(r_\mathrm{cut}/2)^2$ with $r_\mathrm{cut}=1$, even though the particles did not interact. Further, we used $D_\mathrm{r}=0.01$ for the rotational diffusion constant and $\Delta t=0.001$ for the time stepping procedure.

We averaged the statistics over $N_\mathrm{seed}=100$ random initial conditions for all plots except for the effective diffusion study, i.e. Fig.~\ref{fig:predicted-and-measured-msd}, Fig.~\ref{fig:msd-scaling}, Fig.~\ref{fig:lt-diff-const}, and Fig.~\ref{fig:msd-trans-noise}, for which we used $N_\mathrm{seed}=20$.

In all simulations that involved the external field, we used $k_\mathrm{r}=0.01$ and $D_\mathrm{c}=1$ for the reaction and diffusion constants. The field was discretized with $N_\mathrm{g}=101$ grid points resulting in a grid spacing of $\Delta=L/(N_\mathrm{g}-1)$. For the concentration-dependent swimming speed \eqref{eq:spp-tanh}, we chose $c_\mathrm{typ}=0.35$ for the typical value of the concentration field,  $v_\mathrm{min}=0.25, v_\mathrm{max}=0.5$, for the minimal and maximal velocity,  and $w=0.05$ for the transition width  between both velocities.

In all simulations with particle-particle interactions, we used a rather 'hard' harmonic potential with $\varepsilon=k/\gamma_\mathrm{t}=100$ so that the particles would not pass through each other. Further, we always set the effective diameter of disks to $r_\mathrm{cut}=1$ and the radii of the beads of the dumbbell to $a_1=0.5$ and $a_2=0.25$ and their distance to $l=0.5$. In the cases with torques, we set $\beta=k/(\gamma_\mathrm{r}\varepsilon)=1$.

Finally, for rocking motion we set the frequency to $f=1$ and the amplitude-offset ratio to $v^\mathrm{amp}_\alpha(c)/v^\mathrm{off}_\alpha(c)=1.75$.

All remaining parameters for each figure in the main text and the appendix are listed below. These are the packing fraction $\phi_0$, the constant self-propulsion speed or offset velocity $v_\alpha(c)=v^\mathrm{off}(c)=\mathrm{const.}=v_0$ in the cases without external field, the distance of the center of mass from the geometic center $d_\mathrm{gc}$, the relative angle between the center of mass and the orientation vector $\varphi_\mathrm{gc}$, and the number of time steps $N_t$.

\begin{figure}[!htb]
\includegraphics[width=\columnwidth]{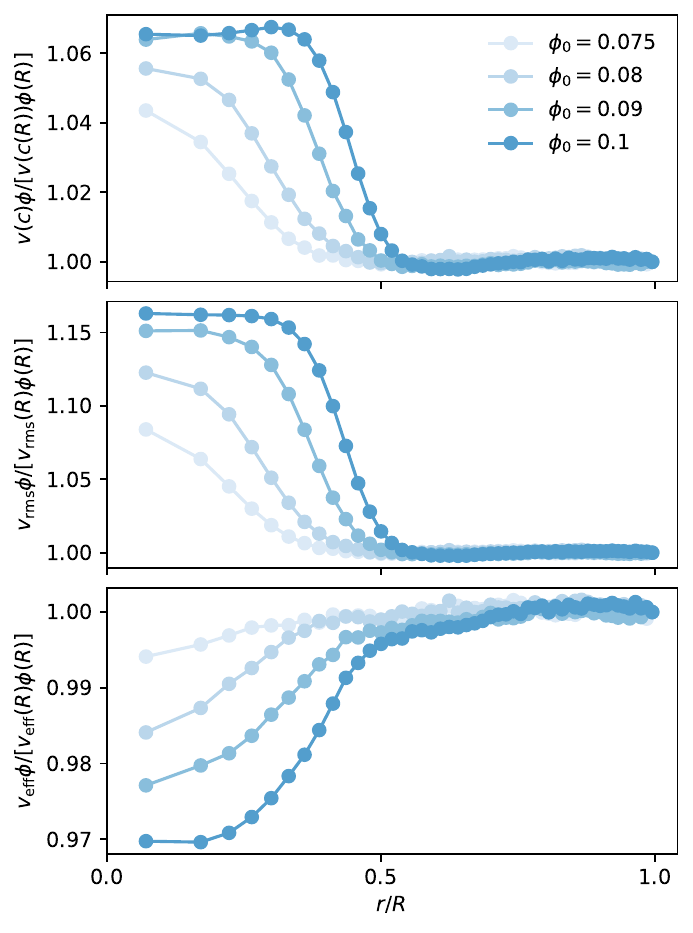}
\caption{The deviations from the inverse scaling behavior decay with decreasing packing fraction as shown here for model IV b ($\varphi_\mathrm{gc}=3\pi/4,d_\mathrm{gc}=0.4$).}
\label{fig:scaling-model-IV-b}
\end{figure}

\begin{itemize}
	\item[-] Fig.~\ref{fig:snapshot-and-radial-profiles} features disks inside an external field (model II, $\phi_0=0.1, N_t=12\cdot10^6$).
	\item[-] Fig.~\ref{fig:scaling} features all models inside an external field (models I, II, III, IV b, IV c, V, $\phi_0=0.1, N_t=12\cdot10^6$, model IV a $\phi_0=0.08, N_t=12\cdot10^6$, model IV a $\varphi_\mathrm{gc}=0, d_\mathrm{gc}=0.4$, IV b $\varphi_\mathrm{gc}=3\pi/4, d_\mathrm{gc}=0.4$, IV c $\varphi_\mathrm{gc}=\pi, d_\mathrm{gc}=0.4$).
	\item[-] Fig.~\ref{fig:force-fields} features disks inside an external field (model II, $\phi_0=0.1, N_t=12\cdot10^6$).
	\item[-] Fig.~\ref{fig:force-torque-contribution} features all models inside an external field but model I (see Fig.~\ref{fig:scaling}).
	\item[-] Fig.~\ref{fig:mips-characterization} features disks, front-heavy disks, and disks with rocking motion without external field (models II*, IV a*, V*, $\phi_0=0.5, v_0=0.5, N_t=12\cdot10^6$, IV a*, $\varphi_\mathrm{gc}=0,d_\mathrm{gc}=0.4$) and a parameter scan $\phi_0=0.1,0.2,0.3,0.4,0.5$.
	\item[-] Fig.~\ref{fig:bands-characterization} features dumbbells and back-heavy disks without external field (models III*, IV c*, $\phi_0=0.3,0.5, v_0=0.5, N_t=12\cdot10^6$, IV c*, $\varphi_\mathrm{gc}=\pi,d_\mathrm{gc}=0.25$) and a parameter scan $\phi_0=0.1,0.2,0.3,0.4,0.5$
	\item[-] Fig.~\ref{fig:clash-of-clusters-characterization} features side-back-heavy disks without external field (model IV b*, $\phi_0=0.3,0.5, v_0=0.25,0.5, N_t=12\cdot10^6, \varphi_\mathrm{gc}=3\pi/4,d_\mathrm{gc}=0.4$) and a parameter scan $v_0 = 0.25,0.5, \phi_0=0.1,0.2,0.3,0.4,0.5$
	\item[-] Fig.~\ref{fig:collective-behavior-with-external-field} features models II to V with external field (models II, III, IV a, IV b, IV c, V, $\phi_0=0.5, N_t=12\cdot10^6$, IV a, $\varphi_\mathrm{gc}=0,d_\mathrm{gc}=0.4$, IV b, $\varphi_\mathrm{gc}=3\pi/4,d_\mathrm{gc}=0.4$, IV c, $\varphi_\mathrm{gc}=\pi,d_\mathrm{gc}=0.4$)
	\item[-] Fig.~\ref{fig:local-packing-fraction} features disks without external field (model II*, $\phi_0=0.5, v_0=0.5, N_t=12\cdot10^6$)
	\item[-] Fig.~\ref{fig:predicted-and-measured-msd} features point-like particles with additional translational noise without an external field (model I with translational noise, $\phi_0=0.15, v_0=0.1, L=100, D_\mathrm{t}=0.01, N_t=10^6$)
	\item[-] Fig.~\ref{fig:msd-scaling} features disks without external field (model II, $\phi_0=0.15, v_0=\sqrt{2}\cdot0.01, L=100, N_t=10^7$)
	\item[-] Fig.~\ref{fig:lt-diff-const} features disks without external field (model II, $L=100, N_t=10^7$, parameter scan $\phi_0=0.05,0.075,0.1,0.2,\dots,0.6, v_0=\sqrt{2}\cdot0.01$, parameter scan $\phi_0=0.15, v_0=0.005, 0.01, 0.02,0.04,\dots,1.28$)
	\item[-] Fig.~\ref{fig:msd-trans-noise} features disks with additional translational noise without an external field (model II with translational noise, $\phi_0=0.15, L=100, N_t=10^7,D_\mathrm{t}=0.01$, first $v_0=0.1$, second $v_0=\sqrt{2}\cdot0.01$, third $v_0=0.002$)
	\item[-] Fig.~\ref{fig:scaling-model-IV-b} features disks with shifted center of mass to the side and back (model IV b, $\phi_0=0.075,0.08,0.09,0.1, N_\mathrm{t}=12\cdot10^6, \varphi_\mathrm{gc}=3\pi/4, d_\mathrm{gc}=0.4$).
\end{itemize}

\section{Observables}
\label{app:observables}
In the following we explain how we computed the local packing fraction and the local and global polar and nematic order parameters, and how we identified the clusters.

\begin{figure*}[!htb]
\includegraphics[width=\textwidth]{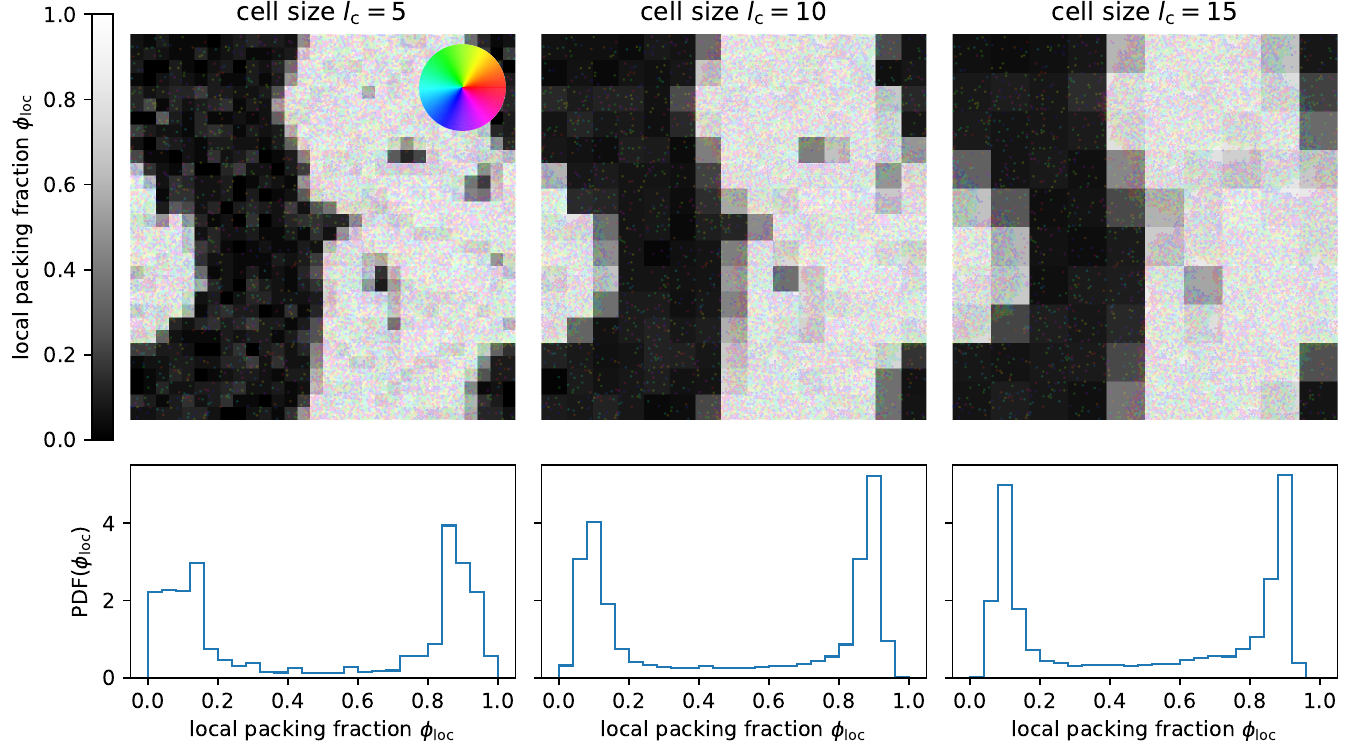}
\caption{
Upper row: Small cells of size $l_\mathrm{c}$ subdivide the simulation domain. The greyscale resembles the local packing fraction $\phi_\mathrm{loc}$ in each cell while the color code depicts the particle orientation for better contrast.
Lower row: The probability density function of the local packing fraction contains the information of all cells over multiple time steps and initial configurations. Its peaks become sharper for larger cell sizes. For the figures in the main text, we chose $l_\mathrm{c}=10$.}
\label{fig:local-packing-fraction}
\end{figure*}

For the local packing fraction and local polar and nematic order parameters, we subdivided the simulation box into smaller cells with area $l_\mathrm{c}^2$, see Fig.~\ref{fig:local-packing-fraction}. We looked at different cell sizes and decided to use an intermediate value so that the particle size is much smaller and the system size much larger than the cell $A_\mathrm{p}\ll l_\mathrm{c}^2\ll A_\mathrm{b}$.

The local packing fraction in cell $j$, i.e.  $\phi_{\mathrm{loc},j}=n_jA_\mathrm{p}/l_\mathrm{c}^2$, is the area fraction of particles inside the cell. The probability density function $\mathrm{PDF}(\phi_\mathrm{loc})$ is the likelihood to find a specific packing fraction, taken over multiple time steps and initial configurations.

The local polar and nematic order parameters are $\psi_{\mathrm{p,loc},j}=|\sum_{k}^{n_j} e^{i\theta_k}|/n_j$ and $\psi_{\mathrm{n,loc},j}=|\sum_{k}^{n_j} e^{2i\theta_k}|/n_j$, respectively~\cite{grossmann2020particle}, where the sum is taken over all particles $k$ inside a cell $j$.

The global polar and nematic order parameters are $\psi_{\mathrm{p,glo}}=|\sum_{k}^{N} e^{i\theta_k}|/N$ and $\psi_{\mathrm{n,glo}}=|\sum_{k}^{N} e^{2i\theta_k}|/N$, respectively~\cite{grossmann2020particle}, where the sum is taken over all particles $N$.

We define a cluster as any single particle or group of particles that overlap with each other directly or indirectly via other particles. To identify a cluster, we loop through all particles and determine whether a particle initiates a new cluster or belongs to an existing one. To check this for each particle, one would in principle need to compute the distances to all other particles in the previously identified clusters. Only if the particle does not belong to any of the previously identified clusters, one can initiate a new cluster. We circumvent this rather inefficient approach by viewing clusters as chains of particles that have a link to a previous and a link to a next particle in the chain. This construct can be realized e.g.~by the concept of two linked lists. Whenever a particle shall be added to a chain, we can simply update its links to the previous and next particles inside that chain. So, how do we determine whether a particle shall be added to a chain? We additionally use the concept of a cell neighbor list that is often used in the context of short-ranged particle-particle interactions~\cite{allen2017computer}. The cell neighbor list efficiently identifies a subset of particles that could be neighbors of a tagged particle. We compute the distances between the tagged particle and a particle identified as possible neighbor. If that distance is smaller than the particles' interaction distance, we connect the chains of both particles (first checking whether both particles are already on the same chain). This approach is faster than the brute-force one mentioned above because it allows to loop for each particle through only a smaller subset of particles.

\section{Derivation of the Fokker-Planck equation}
\label{app:derivation-FPE}
A standard way to obtain the Fokker-Planck equation is to define the probability distribution function of the position and orientation of a single particle
\begin{align*}
	\psi(\boldsymbol x, \theta, t) = \langle \delta(\boldsymbol x_\alpha-\boldsymbol x)\delta(\theta_\alpha - \theta)\rangle
\end{align*}
and to take its time derivative~\cite{sevilla2015smoluchowski}
\begin{align*}
	\partial_t \psi &= \langle \dot{\boldsymbol x}_\alpha \cdot \nabla_{\boldsymbol y}\delta(\boldsymbol y)|_{\boldsymbol y = \boldsymbol x_\alpha - \boldsymbol x} \delta(\theta_\alpha-\theta)\rangle\\
	&+ \langle \delta(\boldsymbol x_\alpha - \boldsymbol x) \dot{\theta}_\alpha\partial_{\vartheta}\delta(\vartheta)|_{\vartheta=\theta_\alpha-\theta}\rangle\,.
\end{align*}
The differential operators acting on the delta functions can be manipulated to depend on the sample space variable only, e.g.
\begin{align*}
	\vartheta &= \theta_\alpha - \theta\quad\Leftrightarrow\quad \theta = \theta_\alpha - \vartheta\\
	\Rightarrow\frac{\partial}{\partial\vartheta} &= \frac{\partial}{\partial\theta}\frac{\partial\theta}{\partial\vartheta} = -\frac{\partial}{\partial\theta}\,. 
\end{align*}
This allows us to pull the differential operator in front of the ensemble average.
\begin{align*}
	\partial_t\psi &= -\nabla_{\boldsymbol x} \cdot \langle \dot{\boldsymbol x}_\alpha \delta(\boldsymbol x_\alpha - \boldsymbol x) \delta(\theta_\alpha-\theta)\rangle\\ 
	&- \partial_{\theta}\langle \delta(\boldsymbol x_\alpha - \boldsymbol x)\dot{\theta}_\alpha\delta(\theta_\alpha-\theta)\rangle
\end{align*}
Before we insert the equations of motion~\eqref{eq:langevin-pos} and~\eqref{eq:langevin-ori}, we can, by definition of the conditional probability, split the terms such that we obtain the one-particle pdf multiplied by a prefactor that still needs to be evaluated.
\begin{align*}
	\partial_t\psi &= -\nabla_{\boldsymbol x} \cdot \big[\langle \dot{\boldsymbol x}_\alpha | \boldsymbol x_\alpha = \boldsymbol x, \theta_\alpha = \theta\rangle \psi\big]\\ 
	&- \partial_{\theta}\big[\langle \dot{\theta}_\alpha| \boldsymbol x_\alpha = \boldsymbol x, \theta_\alpha = \theta\rangle\psi\big]
\end{align*}
Now, when we insert the equations of motion~\eqref{eq:langevin-pos} and~\eqref{eq:langevin-ori}, some of the conditional averages can be evaluated analytically, while others need approximations based on physical understanding or numerical simulations. In fact, the terms in the Langevin equation that depend only on properties of one particle can be directly evaluated.
\begin{align*}
	\langle v_\alpha(c)\boldsymbol p_\alpha | \boldsymbol x_\alpha = \boldsymbol x, \theta_\alpha = \theta\rangle &= v(c)\boldsymbol p\\
	\langle \sqrt{2D_\mathrm{r}}\xi_\alpha | \boldsymbol x_\alpha = \boldsymbol x, \theta_\alpha = \theta\rangle\psi &= -D_\mathrm{r}\partial_\theta\psi\,,
\end{align*}
where we used Gaussian integration by parts for the rotational noise term~\cite{novikov1965functionals,sevilla2015smoluchowski}. The remaining terms, which depend on the properties of more than one particle need an approximation or numerical evaluation:
\begin{align*}
	\bm f(\bm x, \theta, t) &= \langle \bm F_\alpha | \bm x_\alpha = \bm x, \theta_\alpha = \theta\rangle\\
	\tau(\bm x, \theta, t) &= \langle T_\alpha | \bm x_\alpha = \bm x, \theta_\alpha = \theta\rangle\,.
\end{align*}
In the main text, we discuss a simulation-driven approximation for these conditional averages. The resulting Fokker-Planck equation is given in~\eqref{eq:fokker-planck}.

\section{Steady state solution and inverse scaling}
\label{app:steady-state-inverse-scaling}

In the following, we derive the steady state solution and the inverse scaling relationship~\cite{schnitzer1993theory,cates2015motility,stenhammar2016light,frangipane2018dynamic,arlt2019dynamics} for non-interacting active disks inside an external field. The equations of motion simplify considerably and correspond to a closed Fokker-Planck equation.
\begin{align}
	\partial_t \psi = -\nabla_{\bm x}\cdot\left[v(c)\bm p\psi\right] + D_\mathrm{r}\partial^2_\theta\psi
\end{align}
We can obtain a steady-state solution if we use the symmetry of the external field. For the radially symmetric system, we can use polar coordinates $r,\varphi$ and write the Fokker-Planck equation as~\cite{malakar2020steady}
\begin{align*}
	\partial_t \psi = &-\cos(\theta-\varphi)\partial_r\left[v(c)\psi\right]\\ 
	&+ \frac{\sin(\theta-\varphi)}{r}\partial_\varphi\left[v(c)\psi\right] + D_r\partial_\theta^2 \psi\,.
\end{align*}
We assume that $\psi$ varies along the external field, i.e. the radial direction, and that it is independent of the azimuthal angle due to the radial symmetry. We also expect no preferential orientation and hence no dependency on the orientation angle. Then, the Fokker-Planck equation reduces in the steady state to
\begin{align*}
	0 &= -\cos(\theta-\varphi)\partial_r\left[v(c)\psi\right]\\
	\Leftrightarrow J(\varphi, \theta) &= \cos(\theta-\varphi)v(c)\psi\,.
\end{align*}
Note that the flux $J(\varphi, \theta)$ can still depend on the azimuthal and the orientation angle. To comply with the assumption that $\psi$ does not explicitly depend on either of the two, the flux has to be proportional to the cosine of the angular difference
\begin{align*}
	J(\varphi, \theta) &= J_0\cos(\theta-\varphi)\\
	\Rightarrow \psi &= \frac{J_0}{v(c)}\,.
\end{align*}
The constant $J_0$ normalizes $\psi$. The particle density is defined as the integral over the orientation angle
\begin{align*}
	\rho(\boldsymbol r) &= \int_0^{2\pi}\mathrm{d}\theta\,\psi(\boldsymbol r, \theta)\\
	\Rightarrow \rho(\boldsymbol r) &\sim v^{-1}(c)\,.
\end{align*}

Note that if we want to compute the radial profile of a pdf analytically, we need to integrate out the azimuthal component.
\begin{align*}
	\rho(r) &= \int_0^{2\pi}\mathrm{d}\varphi\,r\rho(\boldsymbol r)\\
	&\sim 2\pi rv^{-1}(c)\,.
\end{align*}
Numerically, we count the number of particles inside shells of same area and divide this by the shell area. Like this, the additional increase with radius as in the radial profile above is neglected.

\section{Discussion on effective velocity and diffusion}
\label{app:eff-vel-diff}
This and other studies~\cite{bialke2013microscopic,stenhammar2013continuum,stenhammar2014phase} find that steric particle-particle interactions, on average, slow down a particle's motion. While it has been proposed that these interactions can also, on average, speed up (parts of) the particle's motion, it remains yet to be shown numerically. In the following sections, we analyse the mean square displacement in detail and demonstrate that indeed steric particle-particle interactions can induce additional motion that is non-zero on average, but that this motion is typically orders of magnitude smaller than the predescribed motion.

In~\cite{speck2015dynamical}~the authors decompose the mean interaction force acting on a particle into two contributions: one in the direction of the particle orientation and one perpendicular to it. While the former is by definition an effective contribution to the self-propulsion velocity, the authors intuitively assume the latter to be an effective contribution to the translational diffusion. At this point, we would like to raise the question when (if at all) steric particle-particle interactions contribute to a diffusion process. We believe that two steps are helpful to answer this question: a separation of directed and undirected motion via projection as has been proposed by~\cite{speck2015dynamical}~as well as an analysis of time scales via the mean square displacement.

In the following, we consider the equations of motion for particles with constant self-propulsion speed $v_0$, translational and rotational diffusion constants $D_\mathrm{t}$ and $D_\mathrm{r}$ and steric particle-particle interactions as in the model II in the main text:
\begin{align}
\label{eq:eom-with-trans-noise}
	\dot{\bm x}_\alpha &= v_0\bm p_\alpha + \sqrt{2D_\mathrm{t}}\bm \xi^\mathrm{t}_\alpha + \frac{1}{\gamma_\mathrm{t}}\bm F_\alpha\\
	\dot{\theta}_\alpha &= \sqrt{2D_\mathrm{r}}\xi^\mathrm{r}_\alpha\,.
\end{align}

For the separation of targeted and untargeted motion, we use the projection of the instantaneous velocity onto the particle orientation and its orthogonal direction:
\begin{align}
	\dot{\bm y}^\parallel_\alpha &\coloneqq (\dot{\bm x}_\alpha \cdot \bm p_\alpha)\bm p_\alpha\\
	\dot{\bm y}^\perp_\alpha &\coloneqq (\dot{\bm x}_\alpha \cdot \bm p^\perp_\alpha)\bm p^\perp_\alpha\,,
\end{align}
where $\bm p^\perp_\alpha=(-\sin\theta_\alpha,\cos\theta_\alpha)^\mathrm{T}$ and $\bm p_\alpha\cdot\bm p^\perp_\alpha=0$. One can understand these to be the velocities of the motion parallel and orthogonal to the particle orientation, but has to keep in mind that the particle orientation changes direction in each time step.

\begin{figure}[!htb]
\centering
\includegraphics[width=\columnwidth]{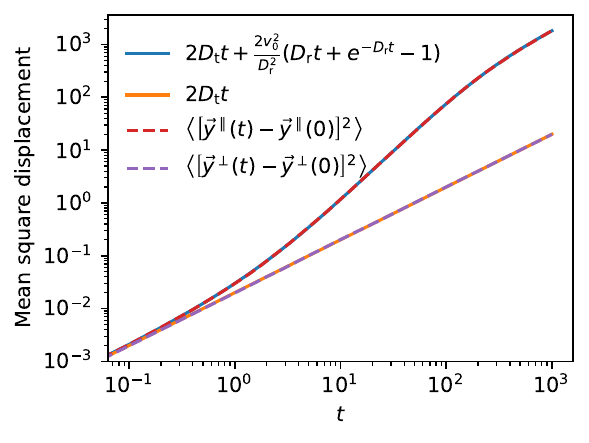}
\caption{Predicted and measured mean square displacement of the motion parallel (blue, red) and orthogonal (orange, purple) to the particle orientation for the system with constant self-propulsion speed, translational and rotational noise but without steric particle-particle interactions.}
\label{fig:predicted-and-measured-msd}
\end{figure}

For the analysis of time scales, we measure the mean-square displacements:
\begin{align}
\mathrm{MSD}(\bm y^\parallel)(t)&=\left\langle\left[ \bm y^\parallel_\alpha(t)-\bm y^\parallel_\alpha(0)\right]^2\right\rangle\\
\mathrm{MSD}(\bm y^\perp)(t)&=\left\langle\left[ \bm y^\perp_\alpha(t)-\bm y^\perp_\alpha(0)\right]^2\right\rangle\,,
\end{align}
for which it is necessary to integrate both quantities over time. Without steric particle-particle interactions the mean square displacements can be evaluated analytically:
\begin{align}
	\mathrm{MSD}(\bm x)(t) &= 4D_\mathrm{t} t + \frac{2v_0^2}{D_\mathrm{r}^2}\left(D_\mathrm{r}t+ e^{-D_\mathrm{r}t}- 1\right)\\
	\mathrm{MSD}(\bm y^\parallel)(t) &= 2D_\mathrm{t} t + \frac{2v_0^2}{D_\mathrm{r}^2}\left(D_\mathrm{r}t+ e^{-D_\mathrm{r}t}- 1\right)\\
	\mathrm{MSD}(\bm y^\perp)(t) &= 2D_\mathrm{t} t\,,
\end{align}
where we averaged over random initial conditions, compare~\cite{ten2011brownian,sevilla2015smoluchowski} and see Fig.~\ref{fig:predicted-and-measured-msd}. Indeed, the motion parallel to the particle orientation has the same mean square displacement as the full motion except for a factor of two in the first term, while the motion orthogonal to the particle orientation is purely diffusive with the same missing factor. The missing factor of two is a consequence of the projection and means that only parts of the translational diffusion occur in one or the other direction, respectively.

With the analytical solution in mind, we postulate that steric particle-particle interactions may contribute to a diffusive motion if their contribution to the mean square displacement of the orthogonal motion is nonzero and (at least over long times) linearly in time. 

\begin{figure}[!htb]
\centering
\includegraphics[width=\columnwidth]{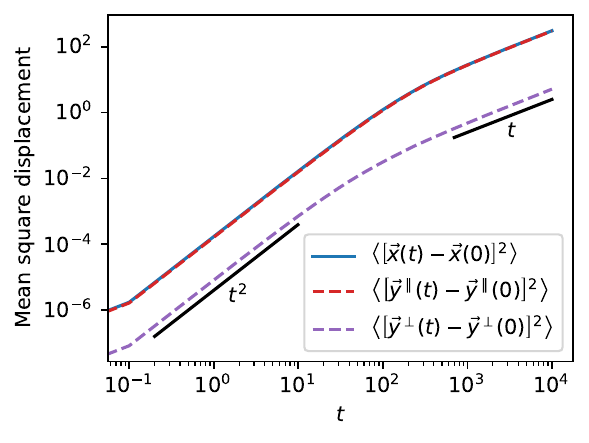}
\caption{Steric particle interactions induce an evasive motion orthogonal to the particle orientation that is ballistic ($\sim t^2$) on small and diffusive ($\sim t$) on large time scales. Here, its mean square displacement is much smaller in magnitude than the one of the parallel motion and contributes only little to the one of the full motion.}
\label{fig:msd-scaling}
\end{figure}

To completely isolate the particle-particle interactions in the orthogonal mean square displacement, we neglect translational diffusion for now. Then we have
\begin{align}
	\dot{\bm y}^\parallel_\alpha &= v_0\bm p_\alpha + \frac{1}{\gamma_\mathrm{t}}\left[\bm F_\alpha\cdot\bm p_\alpha\right]\bm p_\alpha\\
	\dot{\bm y}^\perp_\alpha &= \frac{1}{\gamma_\mathrm{t}}\left[\bm F_\alpha\cdot\bm p^\perp_\alpha\right]\bm p^\perp_\alpha\,,
\end{align}
where the orthogonal motion (second equation) originates from particle interactions only. We find that there is a finite contribution to the orthogonal motion that arises from the steric particle interactions, see Fig.~\ref{fig:msd-scaling}. Its mean square displacement is ballistic on short and diffusive on long time scales. This is equally true for the parallel motion so we may, in the following, identify long-time diffusion constants $D^\parallel_\mathrm{lt}$ and $D^\perp_\mathrm{lt}$. Note, that in this particluar case the mean square displacement of the orthogonal motion is much smaller than the one of the parallel motion.

\begin{figure}[!htb]
\centering
\includegraphics[width=\columnwidth]{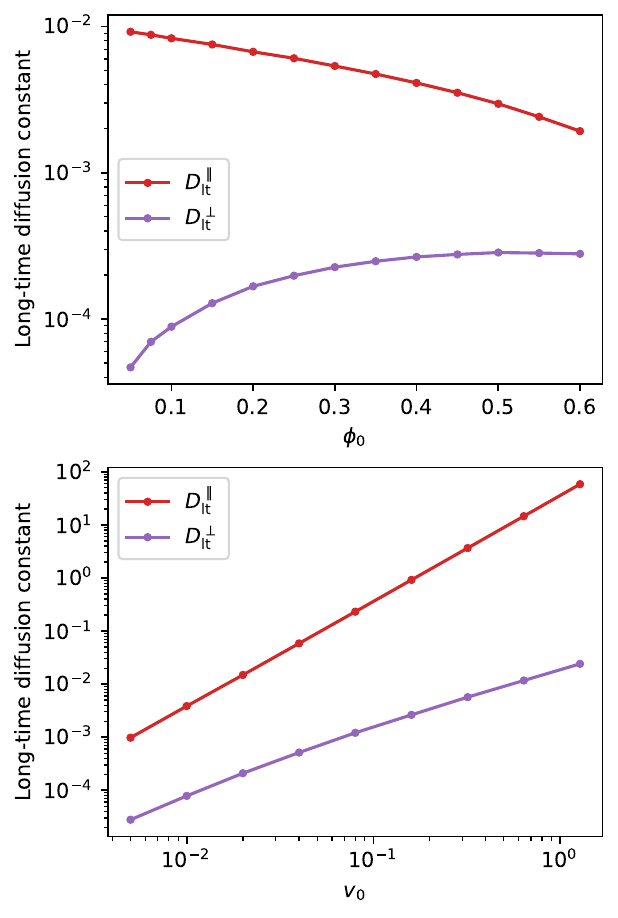}
\caption{The orthogonal long-time diffusion constant $D^\perp_\mathrm{lt}$ is one to multiple orders of magnitude smaller than the parallel long-time diffusion constant $D^\parallel_\mathrm{lt}$. While the former increases for more crowded systems, it seems to reach a plateau for high packing fractions.}
\label{fig:lt-diff-const}
\end{figure}

Next, we investigate how both long-time diffusion constants depend on the global packing fraction $\phi_0$ and the self-propulsion speed $v_0$, see Fig.~\ref{fig:lt-diff-const}. The long-time diffusion constant of the orthogonal motion increases with packing fraction due to an increase in particle interactions but seems to reach a plateau for higher packing fractions. On the opposite, the long-time diffusion constant of the parallel motion decreases with increasing packing fraction which makes intuitively sense, since more particles block the forward motion. Most notably, the difference between the two is one to two orders of magnitude for low to intermediate packing fractions. In contrast, both long-time diffusion constants increase with self-propulsion speed - the parallel one faster than the orthogonal one. Again, the difference of both is on the scale of one to multiple orders of magnitude. We conclude from this that steric particle interactions do induce noise-like motion orthogonal to the particle's orientation, but that this motion is small in comparison to the parallel motion.

\begin{figure}[!htb]
\centering
\includegraphics[width=\columnwidth]{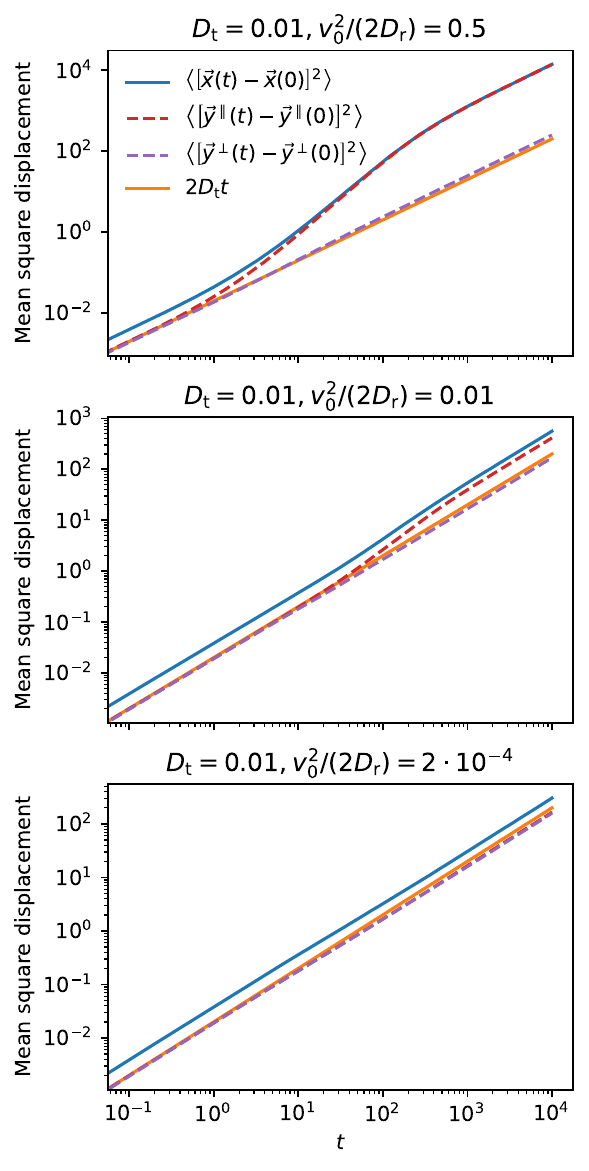}
\caption{The three corner cases: rotational noise (and self-propulsion) dominates, rotational noise and translational noise contribute equally, and translational noise dominates.}
\label{fig:msd-trans-noise}
\end{figure}

Finally, we study the system with additional translational noise, see Eq.~\eqref{eq:eom-with-trans-noise}. One might think according to previous findings that the particle interactions would, at least slightly, increase the orthogonal motion originating from the translational noise. To shed light onto this, we consider three limiting cases that we believe to fully span the range of possible outcomes, see Fig.~\ref{fig:msd-trans-noise}. We use the long-time diffusion constant from the system without interactions
\begin{align}
	D_\mathrm{lt} &= \lim_{t\to\infty}\frac{1}{4t}\text{MSD}(\bm x)(t)\\
	&= D_\mathrm{t} + \frac{v_0^2}{2D_\mathrm{r}}
\end{align}
to distinguish the parameter regime, in which rotational noise dominates the long-time diffusive motion $v_0^2/(2D_\mathrm{r})\gg D_\mathrm{t}$, from the regime, in which rotational and translational contribute equally $v_0^2/(2D_\mathrm{r}) = D_\mathrm{t}$, and the regime, in which translational noise dominates $v_0^2/(2D_\mathrm{r})\ll D_\mathrm{t}$. In the first, we indeed observe a slight increase in the mean square displacement of the orthogonal motion with respect to the mean square displacement for non-interacting particles $2D_\mathrm{t}t$. However, since rotational diffusion and hence parallel motion dominates, the orthogonal motion on long time scales is negligible. In the second and third scenario, we even observe a decrease in the orthogonal motion. We argue that the orthogonal motion due to the translational noise is dampened in magnitude due to particle collisions and that this dampening is stronger than the evasive motion measured without translational noise.

We conclude that steric particle interactions do introduce a diffusion process, if translational noise is absent, and that this diffusion process is small in comparison to the self-propulsion. If, additionally, translational noise is present, the noise is either small enough that steric interactions can (slightly) enhance the diffusion process or it is large enough that steric interactions dampen the diffusion process rather than enhancing it.

We note at this point that in~\cite{van2019interparticle}~in the supplemental material, a study has been performed that measured the change in translational diffusion for passive particles due to interactions and found them negligibly small in comparison to the contribution from self-propulsion if the Péclet number is of order one or larger. This is an additional point for numerically neglecting changes in the translational diffusion.

This study has shown that steric interactions can, in principle, contribute to a diffusion process and that this contribution may be negligible.

\section{Model equations of motion}
\label{app:model-eom}
In this section, we revisit the equations of motion of the five models discussed in the main text. We provide the full equations of model I and based on that all additional terms of the other models. We denote models with an asterisk, e.g. model I*, if we consider the model without the external field, i.e. the self-propulsion speed is constant $v_\alpha(c)=v_0,\forall\alpha$. 

For all models I-V, the external field follows the reaction diffusion equation given by~\citep{fragkopoulos2021self}
\begin{align}
	\partial_t c(\bm r, t) = D_\mathrm{c}\Delta c(\bm r, t) - k_\mathrm{r}\rho(\bm r, t)c(\bm r, t)\,,
\end{align}
where $D_\mathrm{c}$ is the diffusion and $k_\mathrm{r}$ the reaction constant.

Numerically, we solve the equation on a square grid of size $A_\mathrm{g} = (N_\mathrm{g}\Delta)^2$, where $N_\mathrm{g}$ is the number of grid points and $\Delta$ the width between two grid points. We approximate the local particle number density $\rho(\bm r, t)=n_\mathrm{p}/\Delta^2$ by a local step function that counts the number of particles that are closest to a grid point. For boundary conditions, we use the special choice of setting the external field to a constant saturation concentration $c_\mathrm{sat}$ outside a circle with radius $L/2$, where $L$ is the system size.
\subsection{Model I - point-like particles}
\begin{figure}
\includegraphics[width=0.8\columnwidth]{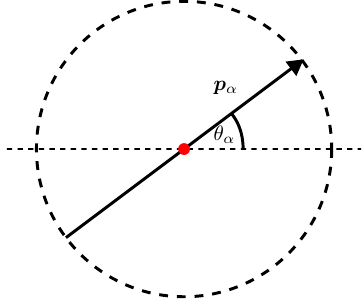}
\caption{The point-like particles have an orientation $\boldsymbol p_\alpha$ that is defined via the orientation angle $\theta_\alpha$ with respect to the horizontal axis. They do not interact with each other as indicated by the dashed circle. The red dot indicates the center.}
\label{fig:point-like-sketch}
\end{figure}

The point-like particles self-propel with speed $v_\alpha(c)$~\cite{fragkopoulos2021self}, which depends on the particles' positions in the external field, in the direction of their orientation vector $\boldsymbol p_\alpha$, see Fig.~\ref{fig:point-like-sketch}.
\begin{align}
	\dot{\bm x}_\alpha &= v_\alpha(c)\bm p_\alpha\\
	\dot{\theta}_\alpha &= \sqrt{2D_\mathrm{r}}\xi_\alpha\\
	\bm p_\alpha &=(\cos\theta_\alpha,\sin\theta_\alpha)^\mathrm{T}\\
	\langle\xi_\alpha(t)\rangle &=0\\
	\langle \xi_\alpha(t)\xi_\beta(t')\rangle &= \delta_{\alpha\beta}\delta(t-t')
\end{align}
Here, $D_\mathrm{r}$ is the rotational diffusion constant and $\xi_\alpha$ a Gaussian white noise that leads to random reorientation of the particle.

\begin{figure}
\includegraphics[width=\columnwidth]{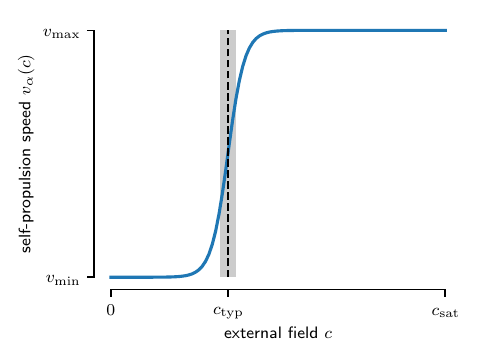}
\caption{The self-propulsion speed of the particles $v_\alpha(c)$ depends on the external field and follows a smoothed step-function. The shaded grey area marks the width $w$ of the transition from low to high speeds.}
\label{fig:spp-vel-vs-conc}
\end{figure}

We interpolate the external field at the particle position bilinearly giving each particle a combination of the external field values of its closest grid points. With that the external field-dependent self-propulsion speed is:
\begin{align}
	v_\alpha(c) = v_\mathrm{min} + &\frac{v_\mathrm{max} - v_\mathrm{min}}{2}\bigg[1+\tanh\bigg(\frac{c(\bm x_\alpha, t)-c_\mathrm{typ}}{w}\bigg)\bigg]\,,
\label{eq-app:spp-tanh}
\end{align}
where the particle can have a minimal and maximal speed of $v_\mathrm{min}$ and $v_\mathrm{max}$, respectively. The transition between those extrema is determined by a 'typical' value $c_\mathrm{typ}$ of the external field and a width $w$, see Fig.~\ref{fig:spp-vel-vs-conc}.


\subsection{Model II - disks}
\begin{figure}
\includegraphics[width=\columnwidth]{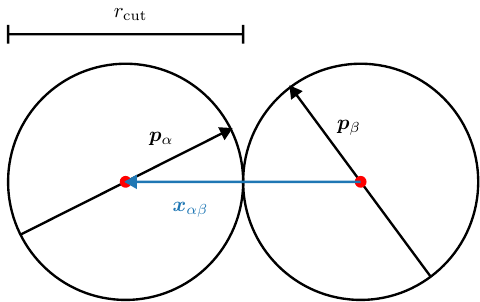}
\caption{The disks interact with each other if the distance of their centers $\boldsymbol x_{\alpha\beta}$ is smaller than their diameter, which is the cut-off radius $r_\mathrm{cut}$ of the harmonic force $\boldsymbol F_\alpha$.}
\label{fig:disk-interaction-sketch}
\end{figure}

\begin{figure}
\includegraphics[width=\columnwidth]{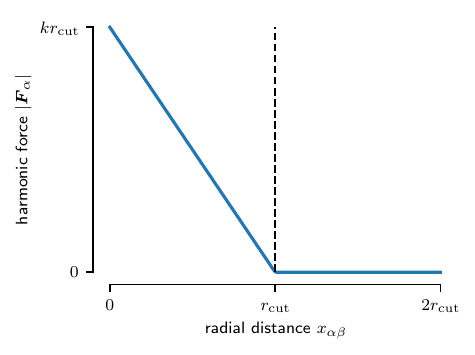}
\caption{The interaction force $\boldsymbol F_\alpha$ decreases linearly with the inter-particle distance and cuts off at the particles' diameter.}
\label{fig:harmonic-potential}
\end{figure}

The active disks additionally interact with each other through a harmonic potential. The potential creates a linear force when the centers of the particles are closer than a cut-off radius $r_\mathrm{cut}$, see Fig.~\ref{fig:harmonic-potential}. The force repels the particles along the distance vector between their centers, see Fig.~\ref{fig:disk-interaction-sketch}.
\begin{align}
	\dot{\bm x}_\alpha &= v_\alpha(c)\bm p_\alpha + \frac{1}{\gamma_\mathrm{t}}\bm F_\alpha\\
	\dot{\theta}_\alpha &= \sqrt{2D_\mathrm{r}}\xi_\alpha\\
	\bm F_\alpha &= k\sum_\beta(r_\mathrm{cut}-x_{\alpha\beta})\hat{\bm x}_{\alpha\beta}\,\quad\text{if }x_{\alpha\beta}<r_\mathrm{cut}\\
	\bm x_{\alpha\beta} &= \bm x_\alpha - \bm x_\beta
\end{align}
Here, $\gamma_\mathrm{t}$ is the translational friction coefficient and $k$ the spring constant of the harmonic potential.

\subsection{Model III - dumbbells}
\begin{figure}
\includegraphics[width=\columnwidth]{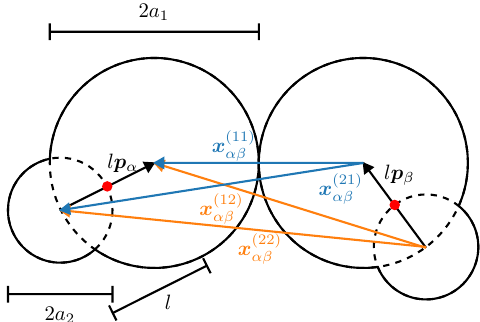}
\caption{The dumbbells consist of a large bead with radius $a_1$ and a small bead with radius $a_2$ which overlap so that the centers of the beads have the distance $l$. The dumbbells interact with each other if one or more of the distances of their beads $\boldsymbol x^{(11)}_{\alpha\beta},\boldsymbol x^{(12)}_{\alpha\beta},\boldsymbol x^{(21)}_{\alpha\beta},\boldsymbol x^{(22)}_{\alpha\beta}$ is smaller than the sum of the radii of the beads that interact, i.e. $2a_1, a_1+a_2, a_2+a_1, 2a_2$ respectively.}
\label{fig:dumbbell-interaction-sketch}
\end{figure}

The dumbbells differ in shape from the disks in that they have two overlapping beads, see Fig.~\ref{fig:dumbbell-interaction-sketch}. The force between interacting dumbbells is the same for each bead as for the disks. The dumbbells have a center of mass that is halfway in between both beads. Hence, we treat the displacement of a single bead as a translation of and rotation about the center of mass. The force acting on the lever between the geometric center of a bead and the center of mass creates a torque that leads to a rotation.
\begin{align}
	\dot{\bm x}_\alpha &= v_\alpha(c)\bm p_\alpha + \frac{1}{\gamma_\mathrm{t}}\sum_{m=1}^2\bm F^{(m)}_\alpha\\
	\dot{\theta}_\alpha &= \sqrt{2D_\mathrm{r}}\xi_\alpha + \frac{1}{\gamma_\mathrm{r}}\sum_{m=1}^2 T^{(m)}_\alpha\\
	\bm F^{(1/2)}_\alpha &= k\sum_\beta\bigg[\bigg(2a_{1/2}-x^{(11/22)}_{\alpha\beta}\bigg)\hat{\bm x}^{(11/22)}_{\alpha\beta}\\
	&+ \bigg(a_{1/2} + a_{2/1}-x^{(12/21)}_{\alpha\beta}\bigg)\hat{\bm x}^{(12/21)}_{\alpha\beta}\bigg]\\
	\bm x_{\alpha\beta}^{(mn)} &= \bm x^{(m)}_\alpha - \bm x^{(n)}_\beta\\
	\bm x_\alpha^{(1/2)} &= \bm x_\alpha \pm l\bm p_\alpha/2\\
	T_\alpha^{(1/2)} &= \left\{\pm l\bm p_\alpha/2\times\bm F_\alpha^{(1/2)}\right\}_z
\end{align}
Here, $\gamma_\mathrm{r}$ is the rotational friction coefficient. $a_1$ and $a_2$ the radii of the larger and smaller bead, respectively, and $l$ the distance between the geometric centers of the beads.
\subsection{Model IV - shifted center of mass}
\begin{figure}
\includegraphics[width=\columnwidth]{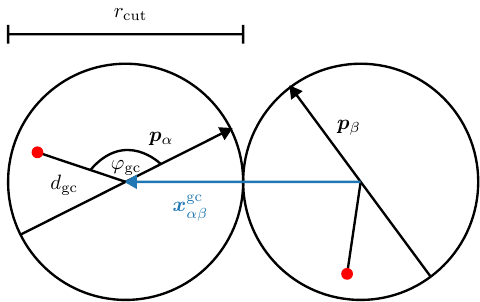}
\caption{The center of mass is shifted with respect to the geometric center of the disk and the orientation vector, defining the distance $d_\mathrm{gc}$ and the relative angle $\varphi_\mathrm{gc}$. The disks interact with each other if the distance of their geometric centers $\boldsymbol x^\mathrm{gc}_{\alpha\beta}$ is smaller than their diameter, which is the cut-off radius $r_\mathrm{cut}$ of the harmonic force $\boldsymbol F_\alpha$.}
\label{fig:shifted-com-interaction-sketch}
\end{figure}

In analogy to the dumbbell model, we can also shift the center of mass independently of the particle shape. Here, we use disks whose center of mass is located at a distance $d_\mathrm{gc}$ from the geometric center and at a relative angle $\varphi_\mathrm{gc}$ with respect to the particle orientation, see Fig.~\ref{fig:shifted-com-interaction-sketch}.
\begin{align}
	\dot{\bm x}_\alpha &= v_\alpha(c)\bm p_\alpha + \frac{1}{\gamma_\mathrm{t}}\bm F_\alpha^\mathrm{gc}\\
	\dot{\theta}_\alpha &= \sqrt{2D_\mathrm{r}}\xi_\alpha + \frac{1}{\gamma_\mathrm{r}} T_\alpha^\mathrm{gc}\\
	\bm x_\alpha &= \bm x^\mathrm{gc}_{\alpha} + d_\mathrm{gc}(\bm p_\alpha\cos\varphi_\mathrm{gc}+\bm p_\alpha^\perp\sin\varphi_\mathrm{gc})\\
	\bm p^\perp_\alpha &=(-\sin\theta_\alpha,\cos\theta_\alpha)^\mathrm{T}\\
	\bm F_\alpha^\mathrm{gc} &= k\sum_\beta(r_\mathrm{cut}-x^\mathrm{gc}_{\alpha\beta})\hat{\bm x}^\mathrm{gc}_{\alpha\beta}\\
	\bm x_{\alpha\beta}^\mathrm{gc} &= \bm x^\mathrm{gc}_\alpha - \bm x^\mathrm{gc}_\beta\\
	T_\alpha^\mathrm{gc} &= \left\{(\boldsymbol x^\mathrm{gc}_{\alpha} - \boldsymbol x_\alpha)\times\boldsymbol F_\alpha^\mathrm{gc}\right\}_z
\end{align}
Like this, we can study the various collective behaviors for different reorientations due to steric interactions.
\subsection{Model V - rocking motion}
Instead of the interaction mechanisms, we can also change the self-propulsion mechanism. Here, the disks interact sterically as in model II and additionally perform a rocking motion which is a essentially a shifted sine function.
\begin{align}
	\dot{\bm x}_\alpha &= v^\mathrm{rm}_{\alpha}(c)\bm p_\alpha + \frac{1}{\gamma_\mathrm{t}}\bm F_\alpha\\
	\dot{\theta}_\alpha &= \sqrt{2D_\mathrm{r}}\xi_\alpha\\
	v^\mathrm{rm}_\alpha(c) &= v^\mathrm{off}_\alpha(c) + v^\mathrm{amp}_\alpha(c)\sin(2\pi ft + \varphi_\alpha)
\end{align}
The offset velocity $v^\mathrm{off}_\alpha(c)$ follows Eq.~\eqref{eq-app:spp-tanh}. The oscillation amplitude $v_\alpha^\mathrm{amp}(c)$ also depends on the external field and we keep the ratio $v_\alpha^\mathrm{amp}(c)/v^\mathrm{off}_\alpha(c)$ constant. The frequency $f$ is independent of the external field. Each particle also starts with a uniformly distributed random phase shift $\varphi_\alpha$.

\begin{figure}
\includegraphics[width=\columnwidth]{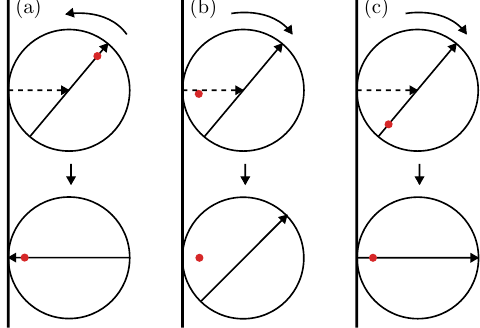}
\caption{Particles that overlap with a wall (or another particle) are subject to a repulsive force (dashed arrow). Here, the particle's geometric center is fixed in position. Hence, the repulsive force causes the particle to rotate until the distance vector between its geometric center and its center of mass (red dot) aligns with the repulsive force. A shifted center of mass is considered in models IV a, b, and c.
}
\label{fig:self-alignment-through-interaction}
\end{figure}

\section{Shifted center of mass}
\label{app:shifted-com}

In a recent review, the role of self-alignment in active matter has been emphasized ~\cite{baconnier2025self}. The basic idea is that for the rigid body of an active particle, various mechanisms can lead to a reorientation that aligns the particle's orientation vector with the force originating from the specific mechanism. In~\cite{baconnier2025self}, the authors discussed two mechanisms in the appendix section: self-alignment due to friction and self-alignment due to a shifted center of mass. 

In the following, we focus on self-alignment due to a shifted center of mass (models IV a, b, c). In this case, it is the repulsive interaction that leads to reorientation of the particle's orientation vector. As an example, consider a particle next to a wall instead of another particle for simplicity, see Fig.~\ref{fig:self-alignment-through-interaction}. Initially, the particle and the wall overlap slightly. As long as they overlap, a repulsive force acts on the geometric center of the particle (dashed arrow). Due to this repulsive force, the particle would be repelled from the wall until there is no further overlap. Since we consider particles with shifted center of mass (red dot), the particle motion could be described by a shift of and a rotation around the particle's center of mass. Imagine now, that the particle's geometric center is fixed. Then, the particle would rotate around the fixed geometric center, until the connecting vector between center of mass and geometric center is aligned with the force acting on the particle. This reorientation is known as self-alignment. Note that in a simulation with moving particles, the particle at the wall (or next to another particle) would in general not stay in place until the alignment has fully taken place. Note also that only for models IV a and c, the final alignment would be with the orientation vector. This is because the center of mass is shifted along the orientation vector in these cases. In model IV b, however, the center of mass is also shifted to the side away from the orientation vector. Then, the interaction force does not align with the orientation vector.

In~\cite[Sec.~II B and App.~A 1]{baconnier2025self}, the authors derive in detail the equations of motion of a rigid body that performs self-propulsion and that is subject to friction and external forces:
\begin{align}
m\ddot{\boldsymbol x}_\alpha &= \boldsymbol F^\mathrm{a}_\alpha - \gamma_\mathrm{t}\dot{\boldsymbol x}^\mathrm{f}_\alpha + \boldsymbol F_\alpha^\mathrm{ext}\\
I\ddot{\theta}_\alpha &= T^\mathrm{a}_\alpha + \{\gamma_\mathrm{t}(\boldsymbol x_\alpha-\boldsymbol x^\mathrm{f}_\alpha)\times\dot{\boldsymbol x}_\alpha\}_z - \gamma_\mathrm{r} \dot{\theta}_\alpha + T^\mathrm{ext}_\alpha\, .
\end{align}
Here, $m$ is the total mass, $\boldsymbol x_\alpha$ the center of mass, $\boldsymbol F_\alpha^\mathrm{a}$ the active self-propulsion force, $\boldsymbol x_\alpha^\mathrm{f}$ the center of friction, $\gamma_\mathrm{t}$ the translational friction coefficient, and $\boldsymbol F_\alpha^\mathrm{ext}$ is the external force; $I$ is the moment of inertia, $\theta_\alpha$ the angle specifying the particle's orientation, $T_\alpha^\mathrm{a}$ the active torque, $\gamma_\mathrm{r}$ the rotational friction coefficient, and $T_\alpha^\mathrm{ext}$ is the external torque. Note, that we already simplified the equations of motion to two dimensions. Also note, that translational and rotational diffusion are neglected here, but can be added to the equations based on phenomenological arguments.

To obtain our equations of motion (model IV), we need to make four assumptions. First, we assume that the center of friction and the center of mass are at the same position, so that $\boldsymbol x_\alpha^\mathrm{f}=\boldsymbol x_\alpha$. Second, we assume that the active self-propulsion force acts directly on the center of mass and not on the geometric center. With this assumption, the active self-propulsion torque is zero and the particle moves (except for rotational noise) in a straight line instead of moving in a circle as a circle swimmer would~\cite{liao2018clustering,liao2021emergent}. Third, we assume that the external force, i.e. steric repulsion between particles, acts on the geometric center so that it leads to a shift of and rotation around the center of mass. Fourth, we take the overdamped limit in both translation and rotation, so that the left-hand side vanishes. The resulting equations of motion are then
\begin{align}
	\dot{\boldsymbol x}_\alpha &= v_0\bm p_\alpha + \frac{1}{\gamma_\mathrm{t}}\bm F_\alpha^\mathrm{gc}\\
	\dot{\theta}_\alpha &= \frac{1}{\gamma_\mathrm{r}}\{(\boldsymbol x^\mathrm{gc}_\alpha - \bm x_\alpha)\times \bm F^\mathrm{gc}_\alpha\}_z\,,
\end{align}
where ``gc" means geometric center and indicates its position and the force acting on it, see model IV in App.~\ref{app:model-eom}.

\bibliography{main}

\end{document}